\documentclass[12pt,onecolumn,spacing=1.5pt]{article}
\usepackage[letterpaper,margin=1in]{geometry}
\usepackage{ulem}

\newcommand{\bd}{\boldsymbol{\textbf{d}}}

\newcommand{\bff}{\textbf{f}}
\newcommand{\bg}{\boldsymbol{g}}

\newcommand{\br}{\boldsymbol{\textbf{r}}}

\newcommand{\bw}{\textbf{w}}

\newcommand{\bB}{\boldsymbol{\textbf{B}}}

\newcommand{\bH}{\boldsymbol{\textbf{H}}}

\newcommand{\bI}{\mathbf{I}}

\newcommand{\bM}{\boldsymbol{\textbf{M}}}
\newcommand{\bL}{\boldsymbol{\textbf{L}}}

\newcommand{\bR}{\boldsymbol{\textbf{R}}}
\newcommand{\bS}{\boldsymbol{\textbf{S}}}

\newcommand{\bX}{\boldsymbol{\textbf{X}}}
\newcommand{\bY}{\boldsymbol{\textbf{Y}}}

\newcommand{\dr}{\,d\br}

\newcommand{\rhod}{\rho_{\scriptsize{\text{data}}}}

\newcommand{\btH}{\widetilde{\boldsymbol{\textbf{H}}}}

\newcommand{\vH}{v_{\rm{H}}}
\newcommand{\vN}{v_{\rm{N}}}

\newcommand{\vxc}{v_{\rm{xc}}}

\newcommand{\Exc}{E_{\rm{xc}}}

\newcommand{\HKSop}{\hat{H}_{\rm{\scriptsize{KS}}}}
\newcommand{\bPsik}{\boldsymbol{\Psi}_k}
\newcommand{\btPsik}{\widetilde{\boldsymbol{\Psi}}_k}
\newcommand{\bpsik}{\boldsymbol{\psi}_k}
\newcommand{\btpsik}{\widetilde{\boldsymbol{\psi}}_k}
\newcommand{\bpsiki}{\boldsymbol{\psi}_k^{(i)}}
\newcommand{\btpsiki}{\widetilde{\boldsymbol{\psi}}_k^{(i)}}
\newcommand{\psik}{\psi_k}
\newcommand{\psiki}{\psi_k^{(i)}}
\newcommand{\psikia}{\psi_{k,a}^{(i)}}
\newcommand{\pki}{p_k^{(i)}}
\newcommand{\pkia}{p_{k,a}^{(i)}}

\newcommand{\bpki}{\textbf{p}_k^{(i)}}
\newcommand{\btpki}{\widetilde{\textbf{p}}_k^{(i)}}
\newcommand{\epsilonk}{\epsilon_k}
\newcommand{\bEk}{\textbf{E}_k}
\newcommand{\trans}{^{\scriptsize{\text{T}}}}
\newcommand{\tr}{\textbf{Tr}}
\newcommand{\lag}{\mathcal{L}}
\newcommand{\bDk}{\mathbf{D}_{k}}
\newcommand{\bPk}{\mathbf{P}_{k}}
\newcommand{\bPsitildek}{\boldsymbol{\widetilde{\Psi}}_{k}}
\newcommand{\bPtildek}{\mathbf{\widetilde{P}}_{k}}
\newcommand{\bQk}{\mathbf{Q}_k}

\newcommand{\epsH}{\epsilon_{\tiny{\text{HOMO}}}}

\newcommand{\modulus}[1]{\left\vert#1\right\vert}

\newcommand{\invDFT}{\texttt{invDFT}~}
\newcommand{\DFTFE}{\texttt{DFT-FE}~}
\usepackage{amsmath}
\usepackage{amssymb}
\usepackage{doi}
\usepackage{mathtools}
\usepackage{hyperref}
\hypersetup{
    colorlinks=true,
    linkcolor=magenta,
    filecolor=magenta,      
    urlcolor=cyan,
    }

\usepackage{subfiles}
\usepackage{graphicx}
\usepackage{graphics}
\usepackage{multirow}
\usepackage{amsxtra}
\usepackage{stmaryrd}
\usepackage{mathrsfs}
\usepackage{caption}
\usepackage{subcaption}
\usepackage{epsfig}
\usepackage{rotating}
\usepackage{setspace}
\usepackage{float}
\usepackage{amsfonts}
\usepackage{amsbsy}
\usepackage{amscd}
\usepackage{amsthm}
\usepackage{relsize}
\usepackage{color}
\usepackage{tablefootnote}
\usepackage{tabularx}
\usepackage{caption}
\usepackage{sidecap}
\usepackage{dcolumn}
\usepackage{footnote}
\usepackage{algorithm}
\usepackage{algorithmic}
\usepackage{breqn}
\usepackage{authblk}

\definecolor{hellgruen}{rgb}{0.2,0.7,0.2}

\newcolumntype{M}[1]{>{\centering\arraybackslash}m{#1}}
\newcolumntype{N}{@{}m{0pt}@{}}
\begin{document}
\title{\invDFT: A CPU-GPU massively parallel tool to find exact exchange-correlation potentials from groundstate densities }
\author[a, \ddag]{Vishal Subramanian}
\author[b,\ddag]{Bikash Kanungo}
\author[a,b, *]{Vikram Gavini}
\affil[a]{\small Department of Materials Science and Engineering, University of Michigan, Ann Arbor, Michigan 48109, USA
\color{black}}
\affil[b]{\small Department of Mechanical Engineering, University of Michigan, Ann Arbor, Michigan 48109, USA }
\affil[$\ddag$]{These authors contributed equally.}
\affil[$*$]{Corresponding author. Email: vikramg@umich.edu}
\date{}
\maketitle
\begin{abstract}
Density functional theory (DFT) remains the most widely used electronic structure method. Although exact in principle, in practice, it relies on approximations to the exchange-correlation (XC) functional, which is known to be a unique functional of the electron density. Despite 50 years of active research, existing XC approximations remain far from general purpose chemical accuracy of various thermochemical and materials properties. In that light, the inverse DFT problem, of finding the exact XC potential corresponding to an accurate groundstate density, offers an insightful tool to understand the nature of the XC functional as well as aid in the development of more accurate functionals. However, solving the inverse DFT problem is fraught with several numerical challenges, such as non-uniqueness or spurious oscillations in the solution and non-convergence. We present \invDFT as an open-source framework to address the outstanding challenges in inverse DFT and computed XC potentials solely from a target density. We do so by use of a systematically convergent finite-element basis and asymptotic corrections to the target density. We also employ several numerical and high-performance computing (HPC) advances that affords both efficiency and parallel scalability, on CPU-GPU hybrid architectures. We demonstrate the accuracy and scalability of \invDFT using accurate full-configuration interaction (FCI) densities as well as model densities, ranging up to 100 electrons and spanning both weakly and strongly correlated molecules.
\end{abstract}


\noindent \textbf{Program Summary}

\begin{small}
\noindent
{\em Program Title:} \invDFT                                          \\
{\em Developer's repository link:} https://github.com/dftfeDevelopers/invDFT  \\
{\em Licensing provisions:} LGPL v2.1  \\
{\em Programming language:} C/C++, CUDA, HIP \\
{\em Supplementary material:} N/A \\
{\em Nature of problem:} Inverse density functional theory \\
{\em Solution method:} \invDFT solves the inverse DFT problem of finding the exact exchange-correlation (XC) potential from a target groundstate density, usually obtained using a quantum many-body method such as configuration interaction (CI).  It recasts the inverse problem as a partial differential equation constrained optimization (PDE-CO). It uses systematically convergent adaptive and higher-order finite-element basis to attain completeness and render the inverse problem well-posed. It also employs asymptotic corrections to the Gaussian or Slater densities from CI calculations to alleviate any numerical artifacts in the resulting XC potentials. To solve the underlying Kohn-Sham eigenvalue problem and the adjoint problem, \invDFT uses Chebyshev filtering and Krylov subspace based algorithms, respectively.\\
{\em Restrictions:} Currently, only the spin-restricted formulation of inverse DFT for non-periodic (atoms, molecules, nanoclusters) is supported. It supports target densities defined in terms of Gaussian and Slater atomic orbitals.

\end{small}

\section{Introduction} \label{sec:intro}
The success and importance of density functional theory (DFT)~\cite{Hohenberg1964, Kohn1965, Becke2014} in computational chemistry and materials science can hardly be overstated. It provides a formally exact reduction of the exponentially scaling many-electron Schr\"odinger equation to an effective single electron problem governed by a simple and tractable quantity---the electron density ($\rho(\br)$). Despite the exactness of the theory, in practice DFT has remained far from exact due to the unavailability of the exact exchange-correlation (XC) functional ($\Exc[\rho]$), which encapsulates the quantum many-electron interactions into a mean-field of the $\rho(\br)$. Even with 50 years of active development~\cite{Cohen2012, Burke2012, Becke2014}, existing XC approximations remain far from general purpose chemical accuracy. While a wide range of theoretical and empirical approaches have been proposed to improve the XC functional, a systematic path for further improvement is lacking. In that light, the study of the XC potential ($\vxc(\br)$), given as the functional derivative of $\Exc[\rho]$ with respect to $\rho(\br)$, offers a promising alternative. Given a density ($\rho$), say from accurate quantum many-body method, one can find its unique XC potential ($\vxc(\br)$) through a procedure called the \textit{inverse} DFT problem ~\cite{Zhao1994, Leeuwen1994, Peirs2003, Wu2003, Jensen2018, Kanungo2019, Shi2021}. In other words, the inverse DFT problem connects DFT to the quantum many-body picture. Thus, the XC potential obtained via inverse DFT offers not just crucial data for conventional and machine-learning based XC approximations,~\cite{Schmidt2019, Zhou2019, Kanungo2024b} but also offers critical insights into the deficiencies of existing model XC functionals.~\cite{Nam2020, Kanungo2021, Kanungo2023, Kanungo2024a, Kaplan2024, Pangeni2025}. 

\noindent Considering the importance of inverse DFT, several efforts have been made to solve it numerically over the past 30 years~\cite{Gorling1992, Wang1993, Zhao1994, Leeuwen1994, Tozer1996, Wu2003, Peirs2003, Jacob2011, Gould2014, Ryabinkin2015, Cuevas2015, Ospadov2017, Jensen2018, Kanungo2019, Shi2021, Stuckrath2021, Shi2022, Gould2023, Aouina2023, Tribedi2023, Kanungo2023}. Broadly, the inverse DFT methods can be classified as either iterative methods~\cite{Gorling1992, Wang1993, Leeuwen1994, Peirs2003, Ryabinkin2012} or constrained optimization methods~\cite{Zhao1994, Wu2003, Jacob2011, Kanungo2019, Kumar2020}.  However, most of the approaches remain prone to numerical artifacts which can be traced back to the incompleteness of the underlying basis (e.g., Gaussian) ~\cite{Burgess2007, Bulat2007,Jacob2011} employed and/or from the finite basis set errors in the the target densities (e.g., Gaussian or Slater densities)~\cite{Mura1997, Schipper1997, Gaiduk2013, Kanungo2019, Kanungo2025} obtained from quantum many-body calculations like configuration interaction (CI). To elaborate, while the continuous problem is well-posed, the discrete problem, if discretized in an incomplete basis, can become ill-posed, leading to non-unique solutions. Further, the lack of correct asymptotics in the target densities also leads to unphysical oscillations in the resulting potential. Gaussian densities, on account of lacking the cusp in the density as well as for incorrect decay in the far-field (Gaussian instead of exponential), are particularly susceptible to spurious oscillations in the potential. Slater densities, which admits a cusp at the nuclei as well as has an exponential decay, alleviates the issue to some extent. However, the Slater densities can still contain basis set errors, especially near the nuclei, and hence, induce artificial oscillations in the potential. Beyond these numerical difficulties, there are two conceptual challenges. First, most of the existing approaches have an implicit assumption of non-degeneracy of the Kohn-Sham eigenvalues, especially for the frontier orbitals, and hence, do not lend themselves applicable to systems with degeneracy. Second, most of the approaches cannot easily handle non-interacting ensemble-v-representable (e-$v_s$) density (i.e., density corresponding to an ensemble of KS determinant) as opposed to non-interacting pure-v-representable (pure-$v_s$ ) density (i.e., density corresponding to a single KS determinant). 

\noindent Given the above outstanding challenges in inverse DFT, the development of software packages for it have been scant. Notably, there have been three attempts at software packages for inverse DFT: \texttt{KS-pies}~\cite{Nam2021}, \texttt{n2v}~\cite{Shi2022}, and \texttt{Serenity}~\cite{Niemeyer2023}. However, all of them employ a Gaussian basis to discretize the inverse DFT problem, which owing to its incompleteness can result in either non-unique solutions and/or unphysical oscillations in the resulting XC potential~\cite{Burgess2007, Jacob2011}. This fundamental difficulty associated with Gaussian basis can limit the reliability of inverse DFT calculations. Additionally, none of the above packages offer any GPU acceleration capability to efficiently harness modern high-performance computing (HPC) architectures. 

\noindent We present \invDFT as on open-source, massively parallel CPU-GPU package to address the various outstanding challenges in inverse DFT as well as offer the accuracy and scalability to realize routine evaluation of accurate XC potentials on modern HPC architectures. \invDFT incorporates our previous formulation for inverse DFT~\cite{Kanungo2019, Kanungo2023} that combines a systematically convergent finite-element (FE) basis along with enforcement of physically relevant asymptotics to the target density and the XC potential to render the inverse DFT problem well-posed. Additionally, it employs various numerical strategies and HPC innovations to attain greater speed and parallel scalability on modern CPU-GPU hybrid architecture. It leverages on the \texttt{DFT-FE}~\cite{MOTAMARRI2020106853, das2022dft} package for certain FE and linear algebra operations. It provides easy interfacing with widely used quantum chemistry codes, such as \texttt{QChem}~\cite{QChem5}, 
\texttt{NWChem}~\cite{NWChem}, \texttt{PySCF}~\cite{PySCF} to feed in accurate target densities from wavefunction-based approaches like configuration interaction (CI) or coupled-cluster (CC) methods. As an initial implementation, \invDFT only supports inverse DFT within the spin-restricted formalism, where we find only one XC potential that yields the target density, as opposed to a spin-unrestricted formalism where two different XC potentials are evaluated for each of the spin-densities. We will extend support for spin-unrestricted inverse DFT in a future release of the package. \invDFT is distributed under LPGL v2.1 and can be accessed from the Github repository~\cite{invDFTGit}.

\section{Theory} \label{sec:theory}
While there are several, in principle, equivalent ways of posing the inverse DFT problem~\cite{Shi2021, Kumar2020}, we solve it as a PDE-constrained optimization. Previously, in the context of spin-unrestricted formalism of DFT, we have proposed a PDE-constrained optimization approach to inverse DFT that can admit degeneracy in Kohn-Sham eigenvalues as well as handle ensemble-v-representable densities~\cite{Kanungo2023}. For completeness, we present the PDE-constrained optimization approach, albeit for the simpler case of spin-restricted DFT as implemented in \invDFT. 

Within the spin-restricted formalism, given a target density ($\rhod(\br)$), the inverse DFT problem of finding its corresponding XC potential ($\vxc(\br)$) can be posed as 
\begin{equation}\label{eq:objective}
\min_{\vxc(\br)}\int w(\br)\left(\rhod(\br)-\rho(\br)\right)^2\dr\,,  
\end{equation}
where  $\rho(\br)$ is the Kohn-Sham (KS) density and $w(\br)$ is a positive weight that expedites the convergence, especially in the low density region. The $\rho(\br)$ is obtained from the solution of the KS eigenvalue problem, which, for a non-periodic system (e.g., atoms, molecules), is given by
\begin{equation}\label{eq:KS}
\HKSop\bPsik(\br)=\bPsik(\br)\bEk\,,\quad k=1,2,\ldots,M\,.
\end{equation}
In the above equation, $\HKSop=-\frac{1}{2}\nabla^2 + \vH(\br) + \vN(\br) +\vxc(\br)$ is the KS Hamiltonian, wherein $\vH(\br)=\int \frac{\rho_{\rm data}(\br')}{|\br-\br'|}\dr'$ is the Hartree potential corresponding to the target density and $\vN(\br) = -\sum_{I}^{L_a} \frac{Z_I}{\modulus{\bR_I - \br}}$ is the nuclear potential corresponding to the $L_a$ nuclei 
with the $I^{\text{th}}$ nucleus of atomic number $Z_I$ located at $\bR_I$. In the above equation, $k$ indexes the distinct eigenvalues, with the $k^{\rm{th}}$ eigenvalue having a multiplicity of $m_{k}$; $\bEk=\epsilonk\bI_{m_{k}}$ is the diagonal eigenvalue matrix with $\epsilonk$ as the $k^{\rm{th}}$ distinct eigenvalue; 
and $\bPsik(\br)=\left[\psik^{(1)}(\br) ~\middle| ~\psik^{(2)}(\br)~ \middle| ~ \cdots ~\middle| ~ \psik^{(m_{k})}(\br)\right]$ comprises of the $m_{k}$ real-valued degenerate eigenfunctions.  While the above formulation is valid for non-periodic systems, the main ideas can also be extended to periodic systems. Typically, one deals with the canonical eigenfunction, which are orthonormal. While the orthogonality between eigenfunctions of different eigenvalues is guaranteed by the Hermiticity of the KS Hamiltonian, orthonormality among degenerate eigenfunctions should be enforced explicitly, i.e., 

\begin{equation} \label{eq:orthonormal}
\int\bPsik\trans(\br)\bPsik(\br)\dr=\bI_{m_{k}}\,.
\end{equation}
Given the canonical eigenfunctions $\{\bPsik\}\rvert_{k=1}^{M}$, the KS density $\rho(\br)$ is defined as
\begin{equation}\label{eq:rho}
    \rho(\br) = \sum_{k=1}^{M}\tr\left(f(\bEk)\bPsik\trans(\br)\bPsik(\br)\right)\,,
\end{equation}
where $f(\bEk)=\left(\bI_{m_{k}}+e^{(\bEk-\mu\bI_{m_{k}})/k_BT}\right)^{-1}$ is Fermi-Dirac occupancy matrix with $\mu$ being the chemical potential given through the conservation of the number of electrons ($N_e$),
\begin{equation} \label{eq:sumf}
    \sum_{k=1}^{M}\tr\left(f(\bEk)\right)=\int \rhod(\br) \dr = N_e\,.
\end{equation}
The above occupancy matrix is crucial to seamlessly handle both pure-$v_s$ and e-$v_s$ densities. 

The optimization of Eq.~\ref{eq:objective} along with the constraint equations can be reformulated as the unconstrained optimization of the following Lagrangian
\begin{equation}\label{eq:L}
\begin{split}
    \lag = & \int w(\br)\left(\rhod(\br)-\rho(\br)\right)^2\dr + \sum_{k=1}^{M}\tr\left(\int\bPk\trans(\br)\left(\HKSop\bPsik(\br)-\bPsik(\br)\bEk\right)\dr\right) + \\
    &\eta\left(\sum_{k=1}^{M}\tr\left(f(\bEk)\right)-N_e\right) 
    + \sum_{k=1}^{M}\tr\left(\bDk\left(\int\bPsik\trans(\br)\bPsik(\br)\dr-\bI_{m_{k}}\right)\right)\,.
\end{split}
\end{equation}
In the above, $\lag$ features three additional quantities that enforce the constraint equations above: (i) $\bPk(\br)=\left[p^{(1)}_{k}(\br) ~\middle| p^{(2)}_{k}(\br)~\middle| \cdots ~\middle|p^{(m_{k})}_{k}(\br)\right]$ are the adjoint functions that enforce KS eigenvalue problem  (Eq.~\ref{eq:KS}) for $\bPsik(\br)$; (ii) $\bDk \in \mathbb{R}^{m_{k}\times m_{k}}$ is the Lagrange multiplier matrix enforcing the orthonormality constraints in Eq.~\ref{eq:orthonormal}; and (iii) $\eta \in \mathbb{R}$ is the Lagrange multiplier enforcing Eq.~\ref{eq:sumf}. It is trivial to show that optimizing $\lag$ with respect to these three additional quantities leads to their corresponding constraint equations. Optimizing $\lag$ with respect to $\bPsik$, $\bEk$, and $\mu$ leads to the following Euler-Lagrange equations, respectively,
\begin{equation}\label{eq:adjoint}
    \HKSop \bPk(\br)-\bPk(\br)\bEk = 4 w(\br)\left(\rhod(\br)-\rho(\br)\right)\bPsik(\br)f(\bEk)
    - \bPsik(\br)\left(\bDk+\bDk\trans\right)\,, 
\end{equation}
\begin{equation}\label{eq:adjointOverlap}
    \int \bPsik\trans(\br)\bPk(\br)\dr = \frac{\partial f_{k}^{\mu}}{\partial\epsilonk}\left[-2 \int w(\br)\left(\rhod(\br)-\rho(\br)\right)\bPsik\trans(\br)\bPsik(\br)\dr+\eta\bI_{m_{k}}\right]\,,
\end{equation}
\begin{equation}\label{eq:eta}
    \eta\sum_{k=1}^{M}m_{k}\frac{\partial f_{k}^{\mu}}{\partial \mu} = 2\sum_{k=1}^{M}\frac{\partial f_{k}^{\mu}}{\partial \mu}\int w(\br)\left(\rhod(\br)-\rho(\br)\right)\tr\left(\bPsik\trans(\br)\bPsik(\br)\right)\dr\,,
\end{equation}
where $f_{k}^{\mu} = \left(1+e^{(\epsilonk-\mu)/k_BT}\right)^{-1}$. 

We note that $(\bDk + \bDk\trans)$ can be evaluated by left multiplying Eq.~\ref{eq:adjoint} with $\bPsik\trans$ and integrating, yielding,
\begin{equation}\label{eq:SI_D}
    \bDk+\bDk\trans=4\left[\int w(\br)\bPsik\trans(\br)\left(\rhod(\br)-\rho(\br)\right)\bPsik\dr\right]f(\bEk)\,,
\end{equation}
where we have used the fact that $\bPsik$ are the eigenvectors of $\HKSop$. Once we solve both the constraint equations (Eq.~\ref{eq:KS}, Eq.~\ref{eq:orthonormal}, and Eq.~\ref{eq:sumf}) as well as the additional Euler-Lagrange equations (Eq.~\ref{eq:adjoint}, Eq.~\ref{eq:adjointOverlap}, and ~\ref{eq:eta}), we can evaluate
\begin{equation}\label{eq:dLvxc}
    \frac{\delta \lag}{\delta \vxc(\br)} = \sum_{k=1}^{M}\tr\left(\bPk\trans(\br) \bPsik(\br)\right)\,.
\end{equation}
The above is the key equation that provides the update to $\vxc(\br)$ via any gradient-based optimization method. A few remarks are in order with regard to degeneracy. In the case of degeneracy, $\bPsik$ is not uniquely defined as any $\bPsitildek=\bPsik\bQk$, where $\bQk$ is an $m_{k} \times m_{k}$ orthogonal matrix, preserves the density (Eq.~\ref{eq:rho}) as well as satisfies KS eigenvalue problem (Eq.~\ref{eq:KS}) and the orthonormality conditions (Eq. ~\ref{eq:orthonormal}). Thus, on the face of it, one has infinitely many choices for the to $\vxc(\br)$ via Eq.~\ref{eq:dLvxc}, rendering the problem ill-posed. However, the choice of our Lagrangian $\lag$ ensures that the adjoint functions corresponding to $\bPsitildek$ are given as $\bPtildek=\bPk\bQk$, which makes Eq.~\ref{eq:dLvxc} invariant to any orthogonal transformation. As a result, the update to $\vxc$ is uniquely determined at each iteration. 
We refer to~\cite{Kanungo2023} for the details on this invariant character of Eq.~\ref{eq:dLvxc}.

\section{Methods} \label{sec:methods}
In this section, we present the various aspects of finite-element based discretization of the continuous equations presented in Sec.~\ref{sec:theory}, efficient strategies to solve the discrete equations, and the various HPC innovations adopted to boost the speed and scalability of the key computational steps involved. 

\subsection{Finite-element based inverse DFT} \label{sec:FE}
\subsubsection{Attaining completeness} \label{sec:complete}
A key reason for numerical artifacts in most of the previous attempts at inverse DFT has been the incompleteness of the basis (e.g., Gaussian) used to discretize the problem. As we have shown in the past~\cite{Kanungo2019, Kanungo2023}, these numerical artifacts can be alleviated by use of a finite-element (FE) basis. Briefly, in the FE basis, the domain of interest is divided into non-overlapping sub-domains called finite-elements. Within each finite-element the basis consists of piecewise polynomial
functions that have a compact support, rendering locality to these basis functions. One can choose from a variety of form and order for the polynomial functions that can be used in constructing the finite element basis. For a comprehensive discourse on finite elements, we refer to~\cite{Hughes2012, Bathe1996}. In this study, we use Lagrange interpolating polynomials. For example, in the 1D case, given a finite-element $e$ with a set of nodes
$\{x_1, x_2, . . . , x_{l+1}\}$ (not necessarily equi-spaced), we can construct $l + 1$ polynomials, each of order $p$, given by
\begin{equation} \label{eq:1Dpoly}
N_{i}^{e,p}(x) = \prod_{\substack{j=1\\ j \neq i}}^{l+1}\frac{x-x_j}{x_i-x_j}\,.
\end{equation}
The 3D basis is constructed by taking tensor products of the 1D Lagrange
polynomials. In the FE basis, one can refine the finite-element size and/or increase the polynomial order to construct a systematically improvable basis. This aspect of the FE basis allows us to attain completeness in the basis, and hence, renders the inverse DFT problem well-posed. We discretize the KS orbitals ($\bpsik^{(i)}(\br)$) and the adjoint functions ($p_k^{(i)}(\br)$) using higher-order FE basis. The XC potential ($\vxc(\br)$), being a much smoother quantity, is discretized using a linear FE basis. 

\begin{figure} 
    \centering
    \includegraphics[scale=0.25]{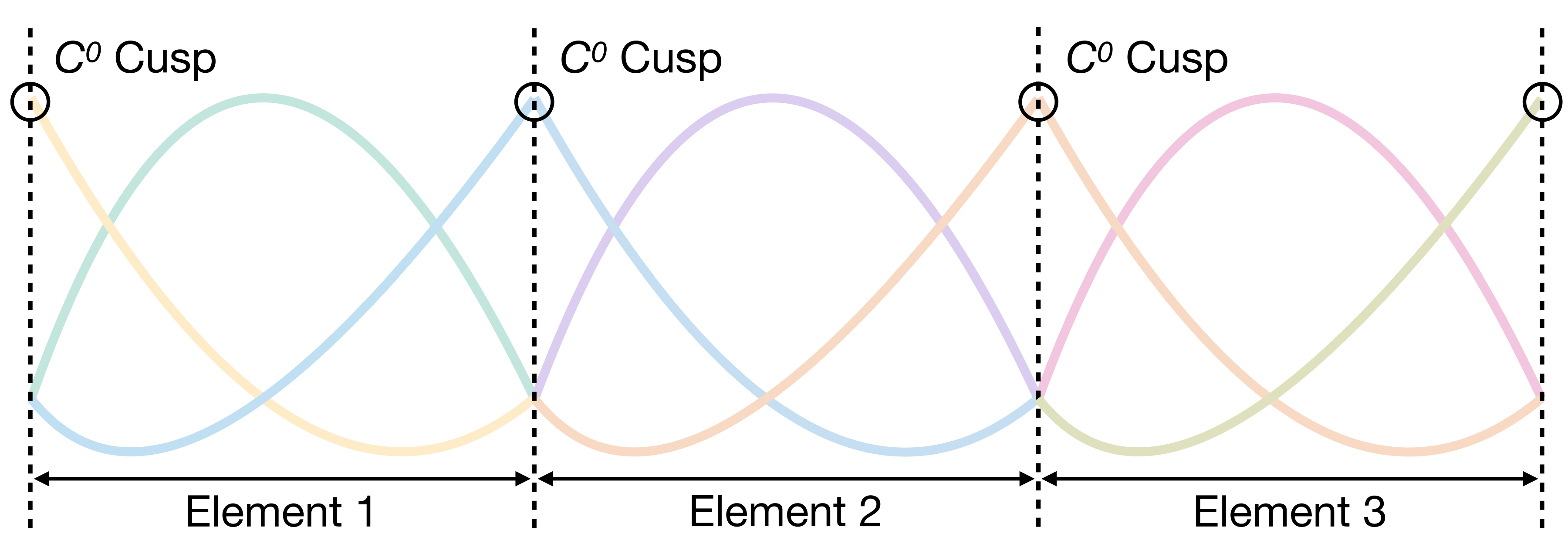}
    \caption{Illustration of three adjacent 1D quadratic finite-elements. The vertical dashed lines denote the boundary between adjacent elements. The black circles highlight
    the $C^0$ continuity (cusp) of the basis at the element boundary.
    }
    \label{fig:FE_1D}
\end{figure}

\noindent We emphasize that, as per the above Lagrange polynomial based construction, the FE basis has
a $C^0$ continuity (cusp) at the element boundary. We illustrate it in Fig.~\ref{fig:FE_1D} for a quadratic finite-element basis. As will be discussed in Sec.~\ref{sec:asymp}, the $C^0$ continuity will be instrumental in alleviating numerical artifacts arising from the Gaussian densities. 

\subsubsection{Ensuring correct asymptotics} \label{sec:asymp}
Besides the completeness of the basis, another major source of numerical artifacts in the solution of inverse DFT problem stemmed from the incorrect asymptotics in the target Gaussian or Slater densities obtained using configuration interaction (CI) or coupled-cluster (CC) calculations. The Gaussian densities lack the cusp at the nuclei~\cite{Kato1957} and have a Gaussian decay (instead of an exponential decay), both of which induce spurious oscillations in the XC potentials obtained via inverse DFT. The Slater densities, by admitting a cusp at the nuclei and having an exponential decay, can remedy some of the deficiencies of Gaussian densities. However, the finite nature of Slater densities can still lead to errors near the nuclei cusp as well as in the far-field, and hence, cause unphysical oscillations in the potential.
In the past, we have have proposed two numerical strategies~\cite{Kanungo2019, Kanungo2025} to alleviate these artifacts, which we incorporate into \invDFT.

\noindent To remedy any oscillations stemming from the incorrect far-field decay in the density, most notably in Gaussian densities, we exploit the known $-1/r$ decay in the exact XC potential. To elaborate, we use an initial guess for XC potential that conforms to the $-1/r$ decay, such as the Fermi-Amaldi potential~\cite{Ayers2005}. Next, during the PDE-constrained optimization, we enforce the update to the XC potential given in Eq.~\ref{eq:dLvxc} to be zero in regions where $\rhod(\br) < \tau_{\text{BC}}$, with $\tau_{\text{BC}} \in [10^{-7}, 10^{-6}]$. In effect, we ensure that the XC potential retains the known $-1/r$ from its initial guess in the low density region, avoiding any artifacts from the incorrect decay of the density. 

\begin{figure}
    \centering
    \includegraphics[scale=0.9]{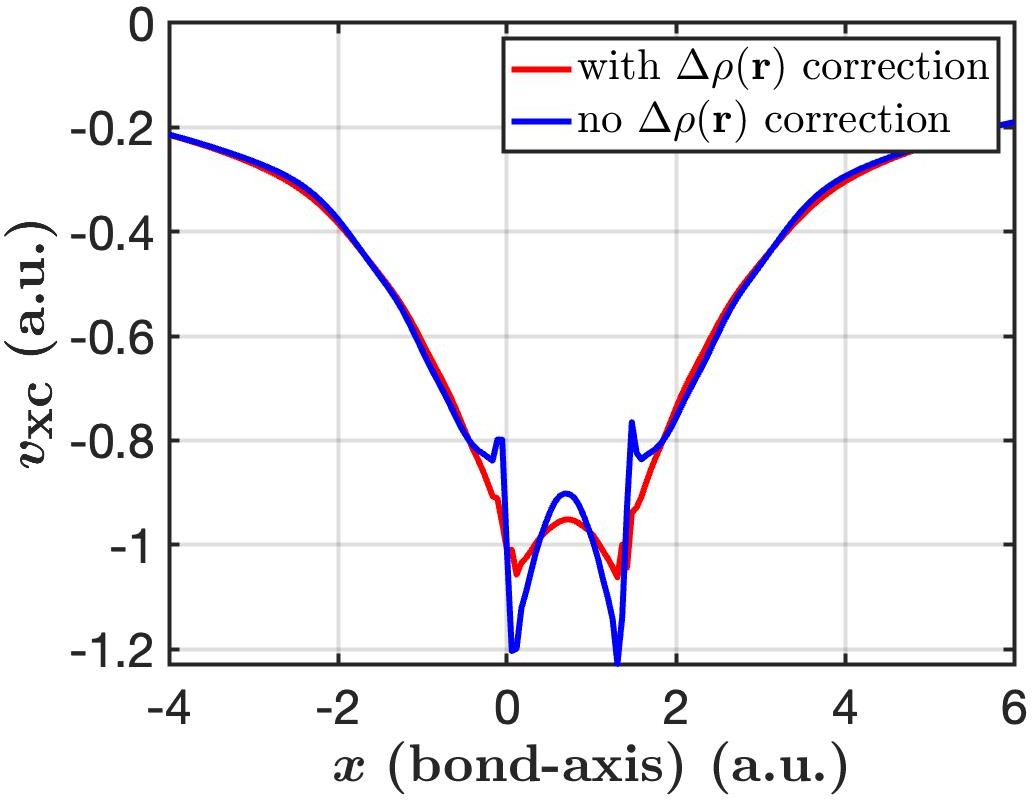}
    \caption{Comparison of XC potential for H$_2$ at equilibrium bond-length with and without the $\Delta\rho$ correction to the Slater density obtained from CI calculations.}
    \label{fig:H2_deltaRho}
\end{figure}

\noindent To remedy the artifacts in the XC potential arising from the basis set errors near the nuclei, we add a small correction ($\Delta \rho(\br)$) to the target density ($\rhod(\br)$), given as
\begin{equation} \label{eq:deltaRho}
    \Delta\rho(\br) = \rho^{\text{DFT}}_{\text{FE}}(\br) - \rho^{\text{DFT}}_{\text{AO}}(\br)\,. 
\end{equation}
In the above, $\rho^{\text{DFT}}_{\text{FE}}(\br)$ is a self-consistent groundstate density for a given density functional (e.g., LDA, GGA) that is solved using an FE basis; and $\rho^{\text{DFT}}_{\text{AO}}(\br)$ is the same density, albeit solved using the Gaussian or Slater atomic orbital basis used in the CI or CC calculation yielding the target density. The key idea is that $\Delta \rho$, being the difference between two different evaluation of
the same physical density (i.e., one with a complete FE basis and the other with an incomplete/finite Gaussian or Slater basis), denotes the basis set error in the Gaussian or Slater density, especially near the nuclei. Moreover, the FE basis, on account of its $C^0$ continuity (see Fig.~\ref{fig:FE_1D}), can also admit a cusp in the density. We note that adding $\Delta \rho(\br)$ to $\rhod(\br)$ does not alter the number of electrons. We ensure a cusp in $\rho^{\text{DFT}}_{\text{FE}}(\br)$ by constructing a finite-element mesh where the nuclei sit on the corner finite-element vertices admitting $C^0$ continuity. The efficacy of the $\Delta \rho$ correction for Gaussian density has been demonstrated in our previous work~\cite{Kanungo2019}. Here, we illustrate the same for Slater density. To do so, we use the H$_2$ molecule at equilibrium bond-length (denoted as H$_2$(eq)) as an example. Figure~\ref{fig:H2_deltaRho} shows the XC potential for H$_2$(eq) corresponding to its CI density~\cite{Kanungo2025} obtained using the QZ4P Slater basis~\cite{Van2003}. 
As evident, the XC potential without the $\Delta \rho$ cusp-correction exhibits large spurious oscillations near the nuclei. These oscillations get alleviated through the use of $\Delta\rho$ correction, resulting in smooth potentials.    

\subsection{Solver Strategies and HPC Considerations} \label{sec:solver}
In this section, we introduce the discrete form of the key equations, the solver strategies employed to solve them, and the various high-performance computing (HPC) advances incorporated to efficiently harness modern CPU-GPU architectures. 
\subsubsection{Discrete equations} \label{sec:discreteEq}
We discretize the various electronic fields, $\psiki$, $\pki$, and $\vxc$, in the FE basis as
\begin{equation} \label{eq:FERep}
    \psi_{k}^{(i),h}(\br) = \sum_{a=1}^{L_1} \psikia N_a^{(l)}(\br); \quad p_{k}^{(i),h}(\br) = \sum_{a=1}^{L_1} \pkia N_a^{(l)}(\br); \quad \vxc^h(\br) = \sum_{b=1}^{L_2} v_{\text{xc},b} N_b^{(1)}(\br)\,. 
\end{equation}
In the above, the superscript $h$ denotes discrete form of the fields; $a$ and $b$ index the finite-element basis with $N_a^{(l)}(\br)$ denoting the $l^{\rm th}$ order compactly supported finite-element basis function. In our calculations, we use the same higher order polynomial---typically third, fourth, or fifth order depending on the highest atomic number in the system---to discretize both the KS orbitals and the adjoint functions. For the XC potential, we use first-order polynomials. Using the above discretization, we get the following discrete form for Eq.~\ref{eq:KS} and Eq.~\ref{eq:adjoint},
\begin{equation}\label{eq:KSDiscrete}
    \bH \bpsiki = \epsilon_k \bM \bpsiki\,, 
\end{equation}
\begin{equation} \label{eq:adjointDiscrete}
    (\bH - \epsilon_k \bM)\bpki = \bg_{k}^{(i)}\,,  
\end{equation}
where $\bpsiki$ and $\bpki$ are the vectors containing the linear coefficients $\psikia$ and $\pkia$ (as defined in Eq.~\ref{eq:FERep}); $\bH$ and $\bM$ are the discrete KS Hamiltonian and the overlap matrix, given as
\begin{equation} \label{eq:HM}
    H_{a,b} = \int N_{a}^{(l)}(\br) \HKSop N_b^{(l)}(\br) \dr, \quad M_{a,b} =  \int N_{a}^{(l)}(\br) N_{b}^{(l)}(\br) \dr\,,
\end{equation}
and $\bg_{k}^{(i)}$ is the right-hand side vector in Eq.~\ref{eq:adjoint} given as,
\begin{equation}
    g_{k,a}^{(i)} = \int \left[4f_k^{\mu} w(\br) \left(\rhod(\br) - \rho(\br)\right)\psiki - \sum_j\left(\bDk + \bDk\trans\right)_{i,j} \psi_k^{(j)} \right] N_a^{(l)}(\br) \dr \,.
\end{equation}
It is apparent that due to the presence of the overlap matrix ($\bM$), the Eq.~\ref{eq:KSDiscrete} and Eq.~\ref{eq:adjointDiscrete} are in the generalized form. Given the availability of more robust and efficient solvers for standard problems, we recast these equations into standard forms, given by
\begin{equation} \label{eq:KSDiscreteStd}
    \btH \btpsiki = \epsilon_k \btpsiki\,,
\end{equation}
\begin{equation} \label{eq:adjointDiscreteStd}
    (\btH - \epsilon_k)\btpki = \widetilde{\textbf{g}}_k^{(i)}\,,
\end{equation}
where $\btH=\bM^{-1/2} \bH \bM^{-1/2}$, $\btpsiki = \bM^{1/2}\bpsiki$, $\btpki=\bM^{1/2}\bpki$, and $\widetilde{\textbf{g}}_k^{(i)}=\bM^{-1/2}\textbf{g}_k^{(i)}$. This transformation from a generalized to standard form requires efficient means to evaluate $\bM^{1/2}$ and $\bM^{-1/2}$. To that end, we use spectral finite-elements along with reduced-order Gauss-Legendre-Lobatto (GLL) quadrature rules. Unlike, the equi-spaced nodes in a standard finite-element, in a spectral finite-element the nodes are distributed according to the GLL points. Thus, when spectral finite-elements are used in conjunction with the GLL quadrature rule, it renders $\bM$ diagonal, greatly simplifying the evaluation of $\bM^{1/2}$ and $\bM^{-1/2}$. In the rest of the paper, unless specified, we simply refer to spectral finite-elements as finite-elements.   

\noindent We solve the inverse DFT problem as a PDE-constrained optimization, which requires iterative update to the XC potential, based on the gradients given by Eq.~\ref{eq:dLvxc}. In a discrete sense, at given iteration $t$, given the vector $\textbf{v}_{\text{xc}}^{(t)}$ containing the linear coefficients $v_{\text{xc},b}$ (as defined in Eq.~\ref{eq:FERep}), the update to the XC potential can be written as
\begin{equation} \label{eq:vxcUpdate}
 \textbf{v}_{\text{xc}}^{(t+1)} = \textbf{v}_{\text{xc}}^{(t)} - \alpha_t \bB_t^{-1} \bff_t\,, 
\end{equation}
where $\alpha_t$ is the step-length, $\bB_t$ is an approximation to the Hessian matrix, i.e.,  $B_{t,ab} = \frac{\delta^2 \lag}{\delta v_{\text{xc},a}^{(t)} \delta v_{\text{xc},b}^{(t)}}$, and $\bff^{(t)}$ is given as
\begin{equation}
    f_b^{(t)} = \sum_{k=1}^M \sum_{i=1}^{m_k} \int \pki(\br) \psiki(\br) N_b^{(1)}(\br) \dr\,.
\end{equation}
We employ a limited-memory Broyden–Fletcher–Goldfarb–Shanno (BFGS)~\cite{Nocedal1980} algorithm to construct approximate $\bB_t^{-1}$ using one-rank updates from a fixed number of previous $\bff_t$.

\begin{figure}
    \centering
    \includegraphics[width=0.9\textwidth]{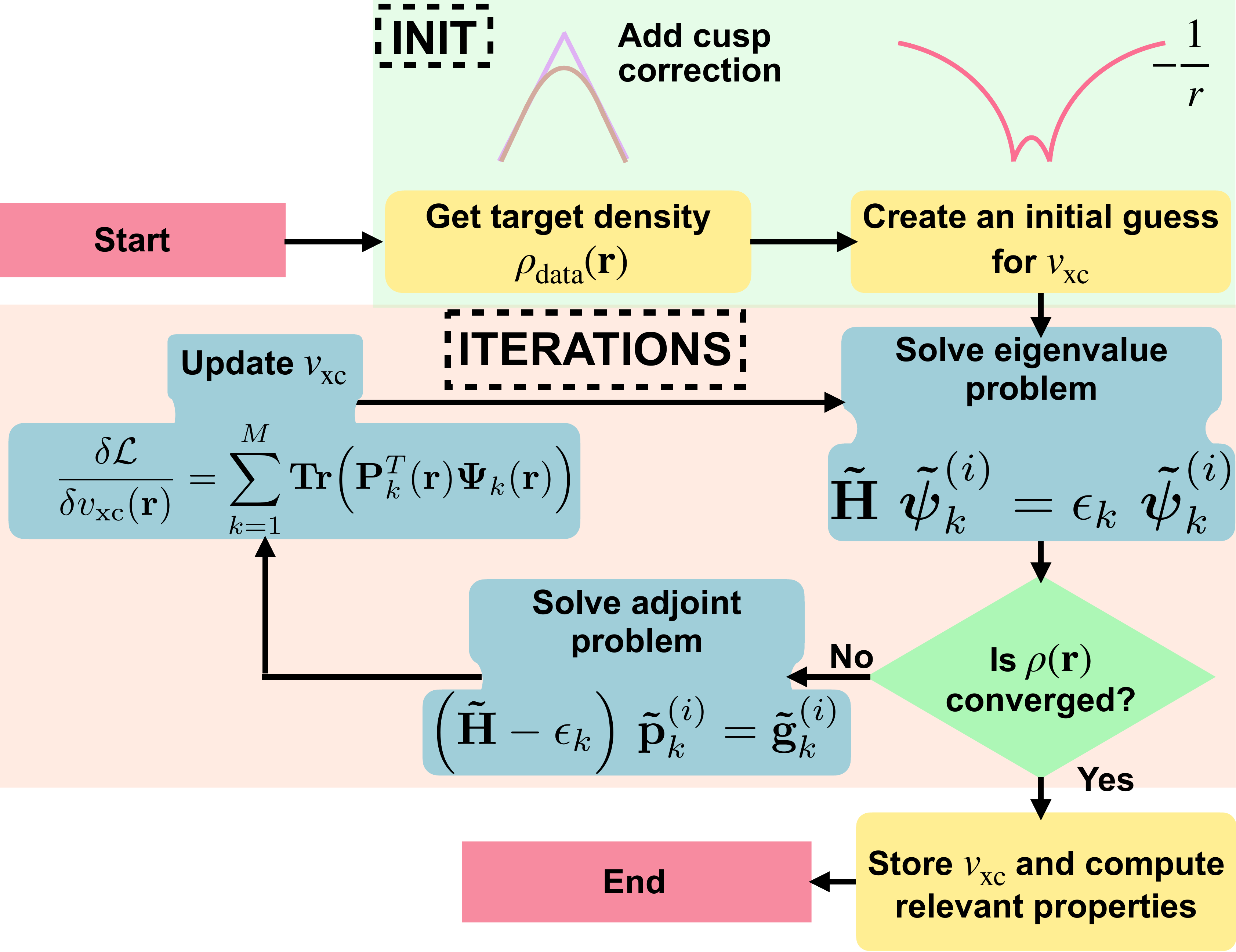}
    \caption{Overview of the inverse DFT algorithm in \invDFT}
    \label{fig:algo_flowchart}
\end{figure}

\noindent We summarize the various steps involved in the inverse DFT algorithm and the discrete equations involved in Fig.~\ref{fig:algo_flowchart}.

 \subsubsection{Efficient Solvers} \label{sec:effsolver}
 Given the XC potential at an iteration $t$, the computationally most dominant steps are the solutions to the Kohn-Sham eigenvalue problem (Eq.~\ref{eq:KSDiscrete}) and the adjoint problem (Eq.~\ref{eq:adjointDiscrete}). We discuss efficient solver strategies implemented in \invDFT to solve them. 
 
 \noindent \textit{Kohn-Sham Eigensolver}: To efficiently solve the KS eigenvalue problem, we exploit the fact that one requires only the low-lying (occupied) eigenpairs of the KS Hamiltonian. To that end, we employ the Chebyshev filtering based subspace iteration (ChFSI) method~\cite{Zhou2006, MOTAMARRI2013308} to create a subspace that closely approximates the occupied eigenspace of $\HKSop$. We refer to the occupied eigenspace of interest to be the `wanted' spectrum and the remainder to be `unwanted' spectrum.  A Chebyshev polynomial of degree $m$, $T_m(x)$, has the following dampening-magnifying property: it takes small values for $x \in [-1,1]$ and grows rapidly outside of it. We consider $\bar{\textbf{H}} = c_1\btH + c_2$, a linear transform that maps the wanted spectrum of $\btH$ to $(-\infty,-1)$ and the unwanted (unoccupied) spectrum to $(-1,1)$. Thus, owing to the dampening-magnifying property of $T_m$, for a given set of input vectors $\bX$, the output vectors $\bY = T_m(\bar{\textbf{H}})\bX$ represent a good approximation for the wanted eigenspace, with the approximation error reducing with increasing $m$. Subsequently, we project the KS eigenvalue problem of Eq.~\ref{eq:KSDiscreteStd} into the space spanned by $\bY$ and solve a problem of much smaller dimension, given as
 \begin{equation} \label{eq:KSProjGen}
     \btH_{\bY} \textbf{u}_k^{(i)} = \epsilon_k \bS_{\bY} \textbf{u}_k^{(i)}\,; \quad \btH_{\bY} = \bY \btH \bY\trans,\quad \bS_{\bY} = \bY\trans\bY\,,
 \end{equation}
 where $\textbf{u}^{(i)}_k$ is the approximation to $\bpsiki$ in $\bY$. We convert the above generalized eigenvalue problem into a standard eigenvalue problem through a Cholesky factorization of $\bS_{\bY} = \bL \bL\trans$, allowing us to rewrite the above as
 \begin{equation}
     \bL^{-1} \btH_{\bY} \left(\bL^{-1}\right)^{\text{T}} \textbf{v}^{(i)}_{k} = \epsilon_k  \textbf{v}^{(i)}_{k}\,;\quad  \textbf{v}^{(i)}_{k} =  \bL\trans\textbf{u}^{(i)}_{k}\,.
 \end{equation}
Upon solving the above eigenvalue problem, we obtain $\bpsiki \approx \bY \left(\bL^{-1}\right)^{\text{T}}\textbf{v}^{(i)}_{k}$. 

\noindent \textit{Adjoint solver}: Solving the adjoint problem (Eq.~\ref{eq:adjointDiscreteStd}) requires some care, as the matrix $(\btH -\epsilon_k \bI)$ in the adjoint problem is singular with $\btPsik=\{\btpsik^{(1)}, \btpsik^{(2)},\ldots,\btpsik^{(m_k)}\}$ as its null vectors. However, the right-hand side of Eq.~\ref{eq:adjointDiscreteStd} is orthogonal to the null vectors, i.e., $\widetilde{\textbf{g}}_{k}^{(i)}$ is orthogonal to $\btPsik$, owing to the KS eigenvalue problem. This ensures at least one solution for the adjoint problem. We note that, given a solution $\bw$ for $\btpki$, $\btpki=\bw + \btPsik \bd$, for any $\bd \in \mathrm{R}^{m_k}$, is also a solution of the adjoint problem. In other words, the component of $\btpki$ along the subspace spanned $\btPsik$ remains undetermined. However, the component of $\btpki$ along $\btPsik$ is known through Eq.~\ref{eq:adjointOverlap}. Using this fact provides a unique solution for $\btpki$.

\noindent In order to efficiently solve the adjoint problem while accounting for the above null space, we employ a preconditioned minimum residual (MINRES)~\cite{Paige1975} solver to solve Eq.~\ref{eq:adjointDiscreteStd}. To elaborate, we use the MINRES method to find  the component of $\btpki$ orthogonal to $\btPsik$ (denoted as $\widetilde{\textbf{p}}_{k,\perp}^{(i)}$) in a Kyrlov space $\mathcal{K}_n=\{\widetilde{\textbf{g}}_{k}^{(i)}, (\btH -\epsilon_k \bI)\widetilde{\textbf{g}}_{k}^{(i)},(\btH -\epsilon_k \bI)^2\widetilde{\textbf{g}}_{k}^{(i)}, \ldots, (\btH -\epsilon_k \bI)^{n-1}\widetilde{\textbf{g}}_{k}^{(i)}\}$. It is straightforward to see that $\mathcal{K}_n$ is orthogonal to $\btPsik$, thereby, guaranteeing that $\widetilde{\textbf{p}}_{k,\perp}^{(i)}$ is orthogonal to $\btPsik$. Further, the component of component of $\widetilde{\textbf{p}}_{k}^{(i)}$ along $\btPsik$ is given by the $i$-th column of the right-hand side of Eq.~\ref{eq:adjointOverlap} (say $\bd_k^{(i)}$). Thus, we have $\btpki=\widetilde{\textbf{p}}_{k,\perp}^{(i)} + \btPsik \bd_k^{(i)}$. In our implementation, we also precondition the MINRES to afford faster convergence. MINRES warrants a positive definite preconditioner, precluding the use of ($\btH - \epsilon_k \bI$), as it is an indefinite matrix. However, as the negative Laplacian, which is a positive definite matrix, is the dominant component of $\btH$, we employ its inverse diagonal as an
inexpensive yet effective preconditioner. 

\subsubsection{HPC Considerations} \label{sec:hpc}
An ongoing trend in modern HPC architectures has been the tremendous increase in single-unit (single CPU/GPU or single node) compute power compared to the slower progress in inter- and intra-node data movement bandwidth. To gainfully utilize modern architectures, the underlying algorithms need to reduce data-movement and data-access while increasing the arithmetic intensity to exploit the fine-grained compute parallelism being offered. Below we discuss the key HPC considerations made in \invDFT to achieve the above. 

\noindent \textit{Improve arithmetic intensity}:  An increasingly common practice in HPC is the batching of vectors, wherever possible, to increase the arithmetic intensity of various linear algebra operations. The two key subspace algorithms in \invDFT---Chebyshev filtering (for Kohn-Sham eigensolve) and the Krylov (for adjoint solve)---are particularly well-suited for batching of vectors, as they involve the solution for multiple Kohn-Sham orbitals or adjoint functions. Thus, in both Chebyshev and Krylov subspace algorithms, we prominently employ batching of vectors to boost the arithmetic intensity.

\noindent \textit{Minimize data-movement and data-access}: To minimize data-access and data-movement costs, we use a memory layout where the coefficients of all the discrete vectors (e.g., discrete KS orbitals, adjoint functions, or any intermediate vectors) belonging to a given FE basis function are contiguously stored. This offers several crucial advantages: (i) on GPUs, it allows for coalesced memory access across GPU threads, boosting the memory bandwidth of reading from the global RAM; (ii) on CPUs, it improves cache locality on many-core architectures; (iii) for message passing interface (MPI) based communication, it allows for simultaneous communication, significantly reducing the network latency.

\begin{figure}[htbp!]
    \centering
    \includegraphics[scale=0.18]{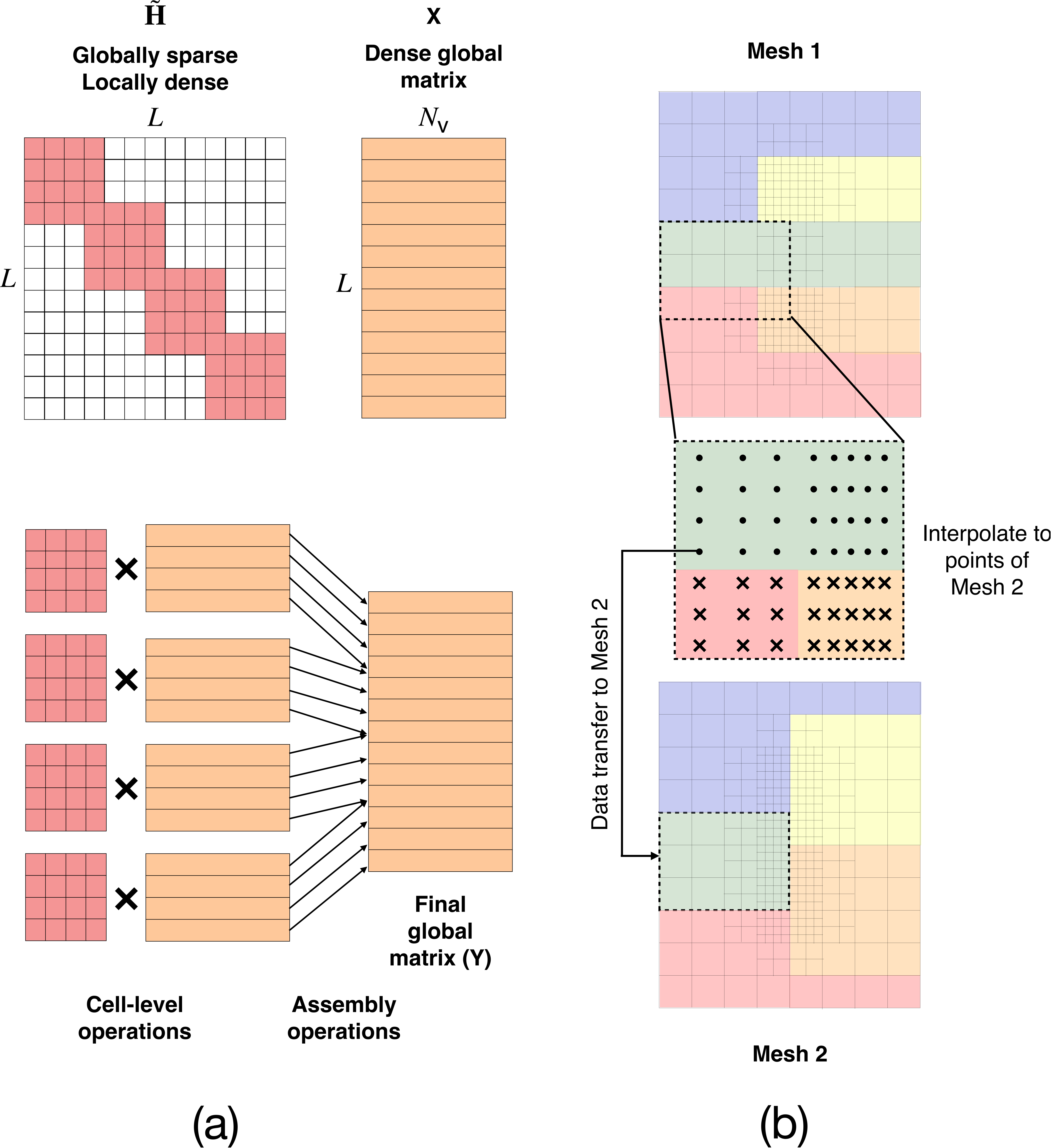}
    \caption{Schematic for the different HPC strategies pursued in this work. (a) Cell-level operations involved in computing  $\btH \bX$. $L$ and $N_v$ denote the number of FE basis and the number of vectors in $\bX$. The top panel shows the typical sparse cell-structure of $\btH$ matrix in an FE basis. This structure allows for decomposing $\bY = \btH \bX$ into cell-level dense-dense matrix operation, followed by an assembly to get $\bY$. $\bX$ has a row-major memory layout: contiguous memory for all $N_v$ vectors for a given FE basis.
    (b) Data transfer between two meshes with incompatible partitioning. The partitioning of mesh 1 and mesh 2 are shown in different colors (color indicate processor index).  Considering the green partition of mesh 2, this regions spans parts of the green, pink, and orange partitions of mesh 1. The middle panel shows the points in that partition where we need the solution field from mesh 1. Circles denote points for which the solution is available locally on mesh 1 (from green partition of mesh 1).  Crosses denote points for which the solution needs to be fetched from other processors on mesh 1 (from pink and orange partition of mesh 1).}
    \label{fig:HPC_scheme}
\end{figure}

\noindent \textit{Fine-grained parallelism}: 
The most critical computational kernel in \invDFT, common to both the Chebyshev filtering (for Kohn-Sham eigensolve) and the Krylov subspace method (for adjoint solve), is the matrix-vector product $\btH\bX$, where $\bX$ generically represents a batch of vectors in the Kohn-Sham eigensolve or the adjoint solve. The $\btH$, being discretized in the FE basis (which is local), is sparse. The $\bX$, in general, are dense set of vectors. Using a global sparse-dense products for $\btH\bX$ involves higher global memory access. However, exploiting the fact that both the $\bH$ and $\bX$ has a local finite-element structure to them, we can recast the sparse-dense matrix products into several
small finite-element dense-dense matrix products. This allows us to perform simultaneous dense-dense matrix products for all the finite-elements in a processor, via vendor optimized batched \texttt{xGEMM} routines. As result, this allows us to harness the fine-grained parallelism in modern CPU-GPU HPC architectures. We refer to Fig.~\ref{fig:HPC_scheme} for a schematic of the above global sparse-dense matrix products to local dense-dense matrix products. We leverage \texttt{DFT-FE}~\cite{MOTAMARRI2020106853, das2022dft, das2019, Das2023} for efficient recasting of the global sparse-dense products into finite-element level dense-dense products. We also utilize the linear algebra engine in \texttt{DFT-FE}, via its API, to perform the batched \texttt{xGEMM} operations.

\noindent \textit{Efficient solution transfer}: As noted earlier, in \invDFT, we use higher-order finite-elements for the KS orbitals ($\bpsiki$) and the adjoint functions ($\bpki$), whereas the XC potential ($\vxc$) is represented using linear finite-elements. This requires efficient transfer of fields across two finite-element meshes. It is customary in finite-elements to attain parallelization via domain decomposition, i.e., assigning different non-overlapping partitions of the mesh to different MPI tasks. Oftentimes, the domain decomposition of the two meshes leads to incompatible partitioning, where the partitions of the two meshes for a given MPI task can differ. This leads to additional difficulties in transferring solution fields between the two meshes. We illustrate this in Fig.~\ref{fig:HPC_scheme}. A naive approach to the solution transfer can lead to $\mathcal{O}(n_t^2)$ computational complexity, where $n_t$ denotes the number of MPI task, affecting the overall scalability of \invDFT. To this end, we have developed an $\mathcal{O}\left(n_t \text{log}(n_t)\right)$ method for field transfer between incompatibly partitioned meshes, based on a combination of tree-based search and sparse point-to-point MPI communication. The details of this method are beyond the scope of this paper, and will be presented in a separate work.

\section{\invDFT Software} \label{sec:software}
We now discuss the various software aspects of \invDFT, ranging from its software architecture, various implementation details, and the user interface. 

\subsection{Overview} \label{sec:invDFTOverview}
\begin{figure}[htbp!]
    \centering
    \includegraphics[width=0.9\textwidth]{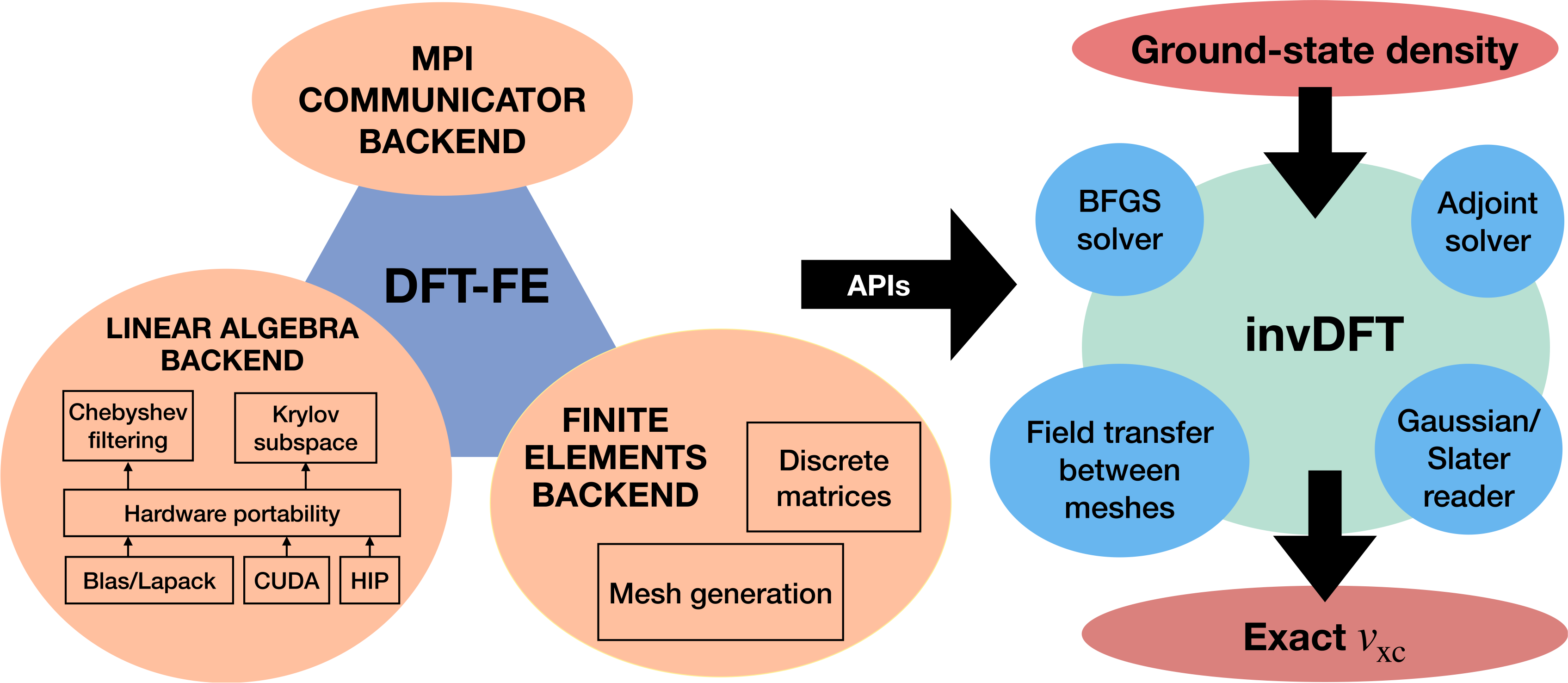}
    \caption{Overview of \invDFT and its various APIs}
    \label{fig:software_schema}
\end{figure}

The software architecture of \invDFT is premised on the fact that the inverse DFT algorithm contains several key pieces that are common to a forward DFT calculation (KS eigenvalue problem). As a result, \invDFT is designed to minimize any duplication of effort by only handling operations that are exclusive to inverse DFT, while leveraging on the capabilities of \DFTFE for operations that are common to forward DFT (see Fig.~\ref{fig:software_schema}). The key components in \invDFT include: (i) density reader to read the target Gaussian or Slater density from a quantum chemistry code; (ii) the BFGS solver to drive the PDE-constrained optimization by iteratively updating the $\vxc$; and (iii) the adjoint solver to solve for the adjoint functions ($\pki(\br)$). The rest of the critical operations are delegated to \DFTFE through its application programming interfaces (APIs). These include the generation of finite-element mesh to discretize the various fields, the construction of the discrete KS Hamiltonian ($\btH$), the action of the discrete matrix on vectors (i.e., $\btH\bX$), and the solution of the KS eigenvalue problem.  Further, we also rely on message passing interface (MPI) and linear algebra (BLAS/LAPACK routines) backends offered by \DFTFE. Particularly, the hardware portability layer in \DFTFE allows \invDFT to seamless execute on both CPU-only and CPU-GPU hybrid architectures, agnostic to the vendor. Currently, the hardware portable layer offers support for only NVIDIA and AMD GPUs, through CUDA and HIP kernels, respectively.      

\subsection{Usage} \label{sec:usage}
We now discuss the main aspects of using the \invDFT package. For a more elaborate explanation of the various input parameters and user choices, we refer to the examples provided in the GitHub repository~\cite{invDFTGit}. We list the important input parameters in Table.~\ref{tab:FEParams}. 

\subsubsection{Target density} \label{sec:targetDensity} 
\invDFT, currently, offers the capability to read Gaussian densities from widely used quantum chemistry packages (e.g., \texttt{QChem}, \texttt{PySCF}, etc.) as the target density. It also offers support to read Slater densities. As noted in Sec.~\ref{sec:asymp}, the lack of correct asymptotics in the Gaussian or Slater densities can induce spurious oscillations in the XC potential. We alleviate by adding a small correction ($\Delta \rho$) to the Gaussian or Slater density (see Eq.~\ref{eq:deltaRho}). Thus, in addition to the target density ($\rhod$), the user also needs to provide the groundstate density using a given density functional ($\rho_{\text{AO}}^{\text{DFT}}$),  solved using the same Gaussian or Slater atomic orbital basis used for $\rhod$. Both $\rhod$ and $\rho_{\text{AO}}^{\text{DFT}}$ can be supplied through their one-particle reduced density matrix (1-RDM) defined in the Gaussian or Slater basis. Further, care must be taken to ensure the same density functional used for $\rho_{\text{AO}}^{\text{DFT}}$ is used to generate $\rho_{\text{FE}}^{\text{DFT}}$ while running \invDFT. Currently, \invDFT supports the use of LDA (PW92~\cite{Perdew1992}) and GGA (PBE~\cite{Perdew1996}) functionals to evaluate $\Delta\rho$. 

\subsubsection{Finite-element basis}\label{sec:meshParams}
\begin{figure}[htbp!]
    \centering
    \includegraphics[width=0.9\textwidth]{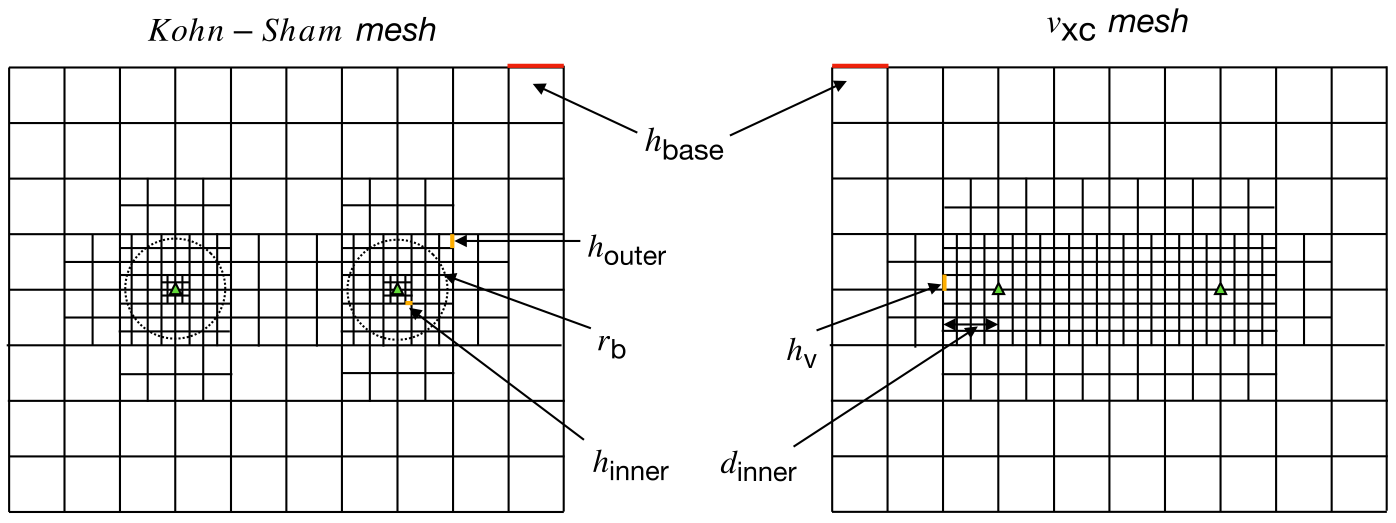}
    \caption{A schematic of the adaptive meshes on which the  Kohn-sham wavefunctions and the $v_{\text{xc}}$ are stored.}
    \label{fig:mesh_schematic_refined}
\end{figure}

A crucial aspect for solving the inverse DFT problem accurately is the choice of the finite-element basis (both element size and polynomial order) used to discretize the various electronic fields involved. At the very outset, the user specifies a cuboidal simulation domain which is discretized using finite-elements. As noted in Sec.~\ref{sec:FE}, we discretize the XC potential ($\vxc$) using linear finite-element. Particularly, we divide the simulation domain into an inner and outer cuboid and use a uniform linear finite-element mesh of size $h_v$ in the inner cuboid. In the outer cuboid, the finite-elements gradually coarsens from $h_v$ to $h_{\text{base}}$, where $h_{\text{base}}$ is automatically determined based on the user-defined simulation domain. The size of the inner cuboid is controlled by the distance $d_{\text{inner}}$ between the inner cuboid's surface and the nearest atom to it. For most cases, we recommend both $h_v$ and $d_{\text{inner}}$ to be unspecified or set to 0, to be heuristically evaluated based on the system. However, if finer control is needed, we recommend the user to vary $h_v$ from 0.1 to 0.02 and $d_{\text{inner}}$ from 5 to 10. To discretize the KS orbitals ($\psi_i$) and the adjoint, we use an adaptive higher-order spectral finite-element (say of order $p$). The adaptivity is controlled by defining a sphere of radius ($r_b$) around each atom and refining the characteristic finite-element size from $h_{\text{outer}}$ to $h_{\text{inner}}$ as one moves from the surface to the center of the sphere. In the far field, the characteristic finite-element size gradually coarsens from $h_{\text{outer}}$ to $h_{\text{base}}$. We illustrate this adaptive meshing scheme in Fig.~\ref{fig:mesh_schematic_refined}. So far, we have discussed four finite-element parameters---$p, r_b, h_{\text{outer}}, h_{\text{inner}}$---to define the adaptive finite-element mesh. While one can easily reduce the discretization errors by increasing $p$ and $r_b$ and/or decreasing $h_{\text{outer}}$ and $h_{\text{inner}}$, careful choices should be made to strike a good balance of accuracy and efficiency. However, making economical choices for four parameters might be non-trivial for a user unfamiliar with the FE basis. We, thus, provide some useful guidelines and defaults. An instructive exercise to determine optimal finite-element parameters is to attain a convergence of 0.1 mHa in energy/atom for an LDA/GGA groundstate calculation. To do so, we recommend the user to start out with $p=4$ and $h_{\text{outer}}=0.5$. We, then, recommend the user to gradually reduce $h_{\text{outer}}$ to 0.1 and/or increase $p$ to 5. For most cases, the other two parameters ($r_b$ and $h_{\text{inner}}$) can be left unspecified or set to 0, for an internal heuristic to pick a close to optimal choices for them. However, if a user needs finer control, we recommend varying $r_b$ from 1.0 to 2.5 and $h_{\text{inner}}$ from 0.1 to 0.01.   

\subsubsection{Initial guess for $\vxc$}\label{sec:vxcInit} 
Given that the inverse DFT problem is a non-linear optimization problem, initial guess for $\vxc$ plays a crucial role in the overall rate of convergence. Currently, \invDFT supports an initial guess of the following form
\begin{equation} \label{eq:init}
    \vxc^{\text{init}}(\br) = f(\br)v_0(\br) [\rhod] + (1-f(\br)) \alpha v_{\text{FA}}[\rhod](\br)\,.
\end{equation}
In the above, $f(\br) = \frac{\rhod(\br)}{\rhod(\br)+\tau}$, with $\tau$ being a small positive number, is a partition function which allows $\vxc^{\text{init}}$ to smoothly transition from $v_0$ to the scaled Fermi-Amaldi potential $v_{\text{FA}}$, as one proceeds from the high density region to low density region ($\rhod <\tau$). The Fermi-Amaldi potential $v_{\text{FA}}[\rhod](\br)=-\frac{1}{N_e}v_{\text{H}}[\rhod](\br)$ guarantees that $\vxc^{\text{init}}$ decays as $-1/r$, as is expected for the XC potential corresponding to a CI or CC density. The scaling factor $\alpha$ allows for a $-\alpha/r$ decay, which can be the expected decay while dealing with densities coming from hybrid XC calculations (e.g., B3LYP, PBE0, SCAN0), wherein $\alpha$ is the fraction of the exact exchange used in the hybrid XC functional. For $v_0$, the initial guess in the high density region, we provide the following form
\begin{equation}
v_0(\br) = v_{\text{x}}^{\text{DFT}}[\rhod](\br) + v_{\text{c}}^{\text{PW92}}[\rhod](\br)\,,    
\end{equation}
where $v_{\text{x}}^{\text{DFT}}$ is either the Slater~\cite{Slater1951} or the LB94~\cite{Leeuwen1994} exchange potential and $v_{\text{c}}^{\text{PW92}}$ is the  PW92~\cite{Perdew1992} correlation potential, all evaluated using $\rhod$. For most cases, we suggest using $\tau=10^{-2}$ and the LB94 exchange potential in $v_0$. 

\subsubsection{Solver parameters}\label{sec:solverParams}
We briefly discuss the key solver parameters involved. We emphasize that, compared to forward DFT, inverse DFT warrants much tighter tolerances in each of the numerical steps involved. The two most important parameters are the tolerances used in the iterative ChFSI-based KS eigensolve and the MINRES-based adjoint solve. We denote $t_{\text{eig}}$ and $t_{\text{ad}}$ to represent the tolerance to which the residual of the eigenvalue problem and the adjoint problem are driven, respectively. Both the values are adaptively tightened as, based on the $L^2$-norm error in the density $e_{L_2}=\sqrt{\int \left(\rhod-\rho\right)^2\dr}$ from the previous iteration. In particular, we set $t_{\text{eig}}=\frac{e_{L_2}^2}{c_{\text{eig}}}$ and $t_{\text{ad}}=\frac{e_{L_2}^2}{c_{\text{ad}}}$. The other crucial parameter is the tolerance $\tau_{\text{BC}}$ for the density, below which the XC potential is kept fixed to its initial guess (to ensure correct far-field decay). The default values for these parameters are $c_{\text{eig}}=10^2$, $c_{\text{ad}}=10^5$, and $\tau_{\text{BC}}=10^{-6}$. 

\begin{table*} [htpb] 
\centering
  \caption{List of important input parameters, their default values, and typical ranges. Default values marked HD and N/A indicate heuristically determined and no default (i.e., must be user-specified), respectively. The arrow direction in the range indicate whether the value should be gradually increased ($\uparrow$) or decreased ($\downarrow$) for convergence.} 
  \label{tab:FEParams}
  \begin{tabular}{| M{2.5cm} | M{7cm} | M{1.5cm}| M{3cm} |}
    \hline
    Parameter & Remark & Default & Typical Range \\ \hline
    $p$ & FE order (Sec.~\ref{sec:meshParams}) & N/A & ${4,5,6,7}$ ($\uparrow$) \\ \hline 
    $h_{\text{outer}}$ & FE mesh parameter (Sec.~\ref{sec:meshParams}) & N/A & $0.1-1.0$ ($\downarrow$) \\ \hline
    $r_b$ & FE mesh parameter (Sec.~\ref{sec:meshParams}) & HD & $1.0-4.0$ ($\uparrow$) \\ \hline
    $h_{\text{inner}}$ & FE mesh parameter (Sec.~\ref{sec:meshParams}) & HD & $0.01-0.1$ ($\downarrow$) \\ \hline
    $h_v$ & FE mesh parameter (Sec.~\ref{sec:meshParams}) & HD & $0.02-0.1$ ($\downarrow$) \\ \hline
    $d_{\text{inner}}$ & FE mesh parameter (Sec.~\ref{sec:meshParams}) & HD & $4.0-10.0$ ($\uparrow$) \\ \hline
    $c_{\text{eig}}$ & Solver parameter (Sec.~\ref{sec:solverParams}) & $10^2$ & $10^2-10^4$ ($\uparrow$) \\ \hline
    $c_{\text{ad}}$ & Solver parameter (Sec.~\ref{sec:solverParams}) & $10^5$ & $10^4-10^7$ ($\uparrow$) \\ \hline
    $\tau_{\text{BC}}$ & Density tolerance (Sec.~\ref{sec:asymp}) & $10^{-6}$ & $10^{-5}-10^{-7}$ ($\downarrow$) \\ \hline
  \end{tabular}
\end{table*}

\section{Results} \label{sec:results}
We present various illustrative examples to demonstrate the accuracy, efficiency, and scalability of the \invDFT package.

\subsection{LDA validation tests} \label{sec:valid}
\begin{figure}
 \begin{subfigure}{0.49\textwidth}
     \includegraphics[width=\textwidth]{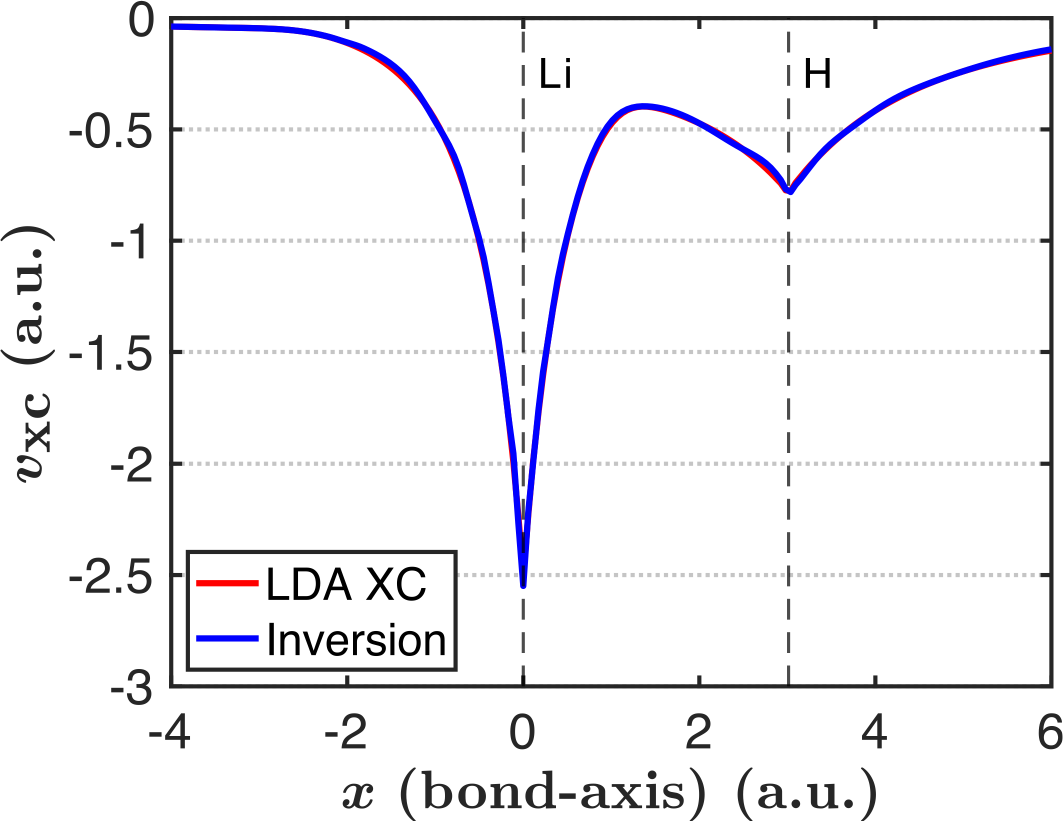}
     \caption{LiH molecule}
     \label{fig:ldaCompPlots_Ne}
 \end{subfigure}
 \hfill
 \begin{subfigure}{0.49\textwidth}
     \includegraphics[width=\textwidth]{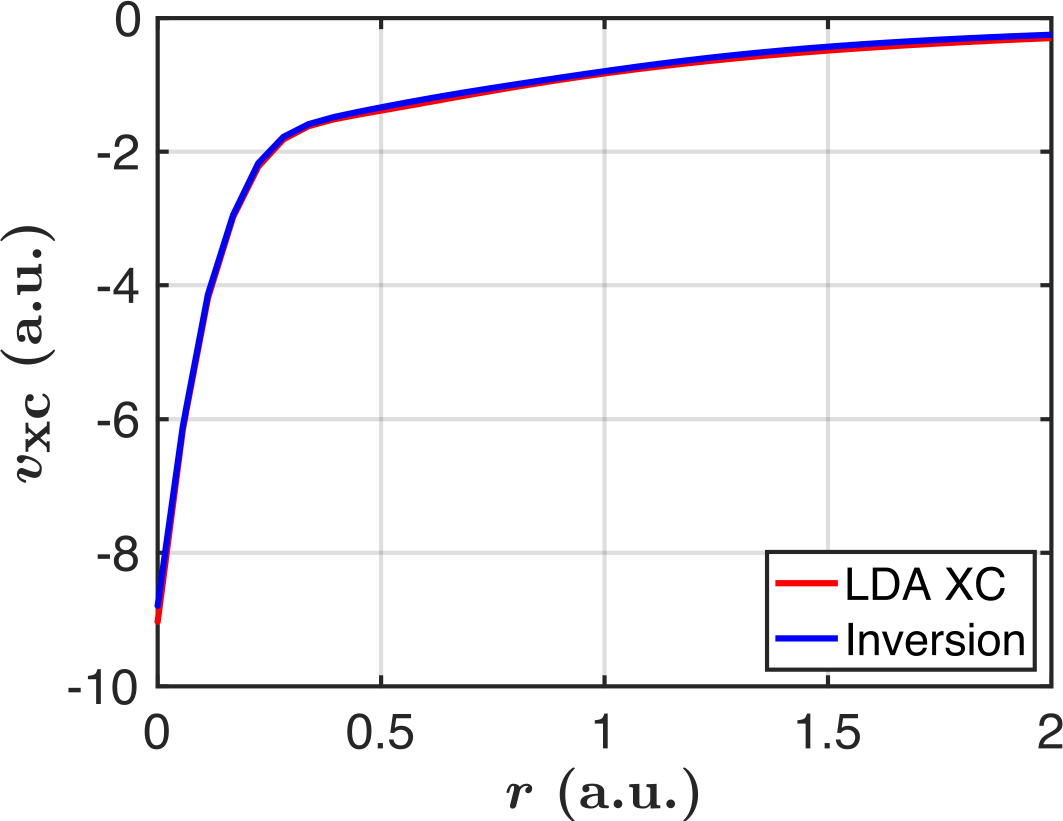}
     \caption{Ne Atom}
     \label{fig:ldaCompPlots_LiH}
 \end{subfigure}

 \caption{Reconstruction of the XC potential corresponding to the LDA density ($\rho_{\text{LDA}}$) for various systems}
 \label{fig:ldaCompPlots}

\end{figure}

We first assess the accuracy of \invDFT through its ability to reconstruct the potential for the groundstate LDA density ($\rho_{\text{LDA}}$), solved using the finite-element basis. Since the corresponding potential, $\vxc^{\text{LDA}}[\rho_{\text{LDA}}]$ is known exactly, we can directly compare the $\vxc$ obtained from an inverse calculation to assess the accuracy of \invDFT. Even though a seemingly straightforward test, many of the earlier attempts which used Gaussian basis suffered from unphysical oscillations and/or non-unique solutions~\cite{Burgess2007, Kananenka2013}, owing to the incompleteness of the basis. Figure~\ref{fig:ldaCompPlots} presents the comparison of the LDA and the reconstructed potential for Ne and LiH. All the calculations are driven to tight accuracy of $||\rho_{\text{LDA}}-\rho||_2 < 10^{-5}$. As evident, the LDA and the reconstructed potentials are virtually indistinguishable. We also attain very close agreement ($< 1$ mHa) between the KS eigenvalues from the LDA and the reconstructed potential, further validating the accuracy of \invDFT.

\subsection{Exact XC potentials from CI densities} \label{sec:exactXC}
We now present the exact XC potentials corresponding to accurate groundstate densities obtained from either incremental full CI (iFCI)~\cite{Zimmerman2017, Zimmerman2017b} or heat bath CI (HBCI)~\cite{Holmes2016, Holmes2017, Sharma2017, Li2018, Dang2022, Chien2018}, both of which are systematic ways to approach the full CI limit in a given basis. Particularly, we reuse the densities that have been used in our previous works~\cite{Kanungo2019, Kanungo2021, Kanungo2023}.
We choose six benchmark systems, namely Ne, LiH, BH, H$_{2}$O, CH$_2$ (singlet), and ortho-C$_6$H$_4$ (ortho-benzyne), that range from weak to strong electronic correlation. To showcase the capability of \invDFT to handle different atomic orbital basis, we consider both Gaussian and Slater basis. For all the systems, we provide the one-electron reduced density matrix (1-RDM), the basis files, and the geometries as Supplementary Materials.  

As noted in Sec.~\ref{sec:asymp}, given the finite basis set error in Gaussian and Slater densities near the nuclei, we add the $\Delta \rho$ correction  to the Gaussian/Slater densities, to alleviate the numerical artifacts stemming from the Gaussian/Slater basis set errors near the nuclei. For all benchmark systems, the inverse problem is deemed converge when $||\rhod-\rho||_2 < 10^{-4}$.  

\begin{figure} [htbp!]
 \begin{subfigure}{0.49\textwidth}
     \includegraphics[width=\textwidth]{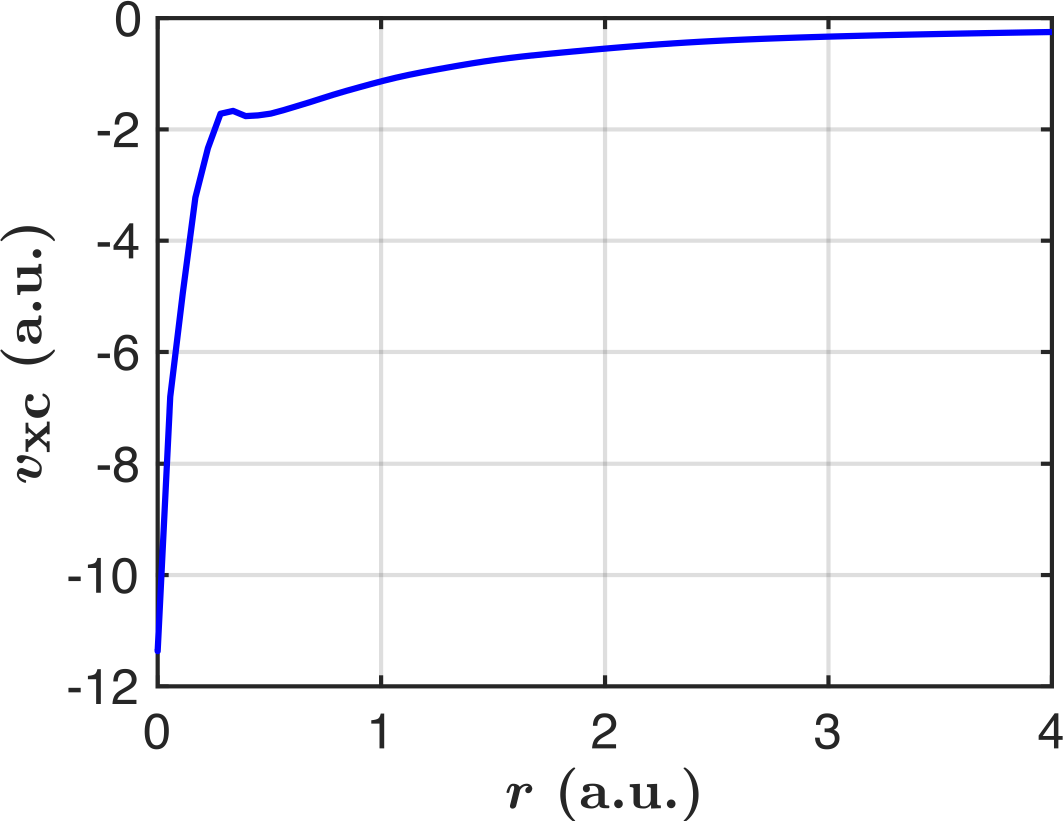}
     \caption{Ne atom}
     \label{fig:Ne}
 \end{subfigure}
 \hfill
 \begin{subfigure}{0.49\textwidth}
     \includegraphics[width=\textwidth]{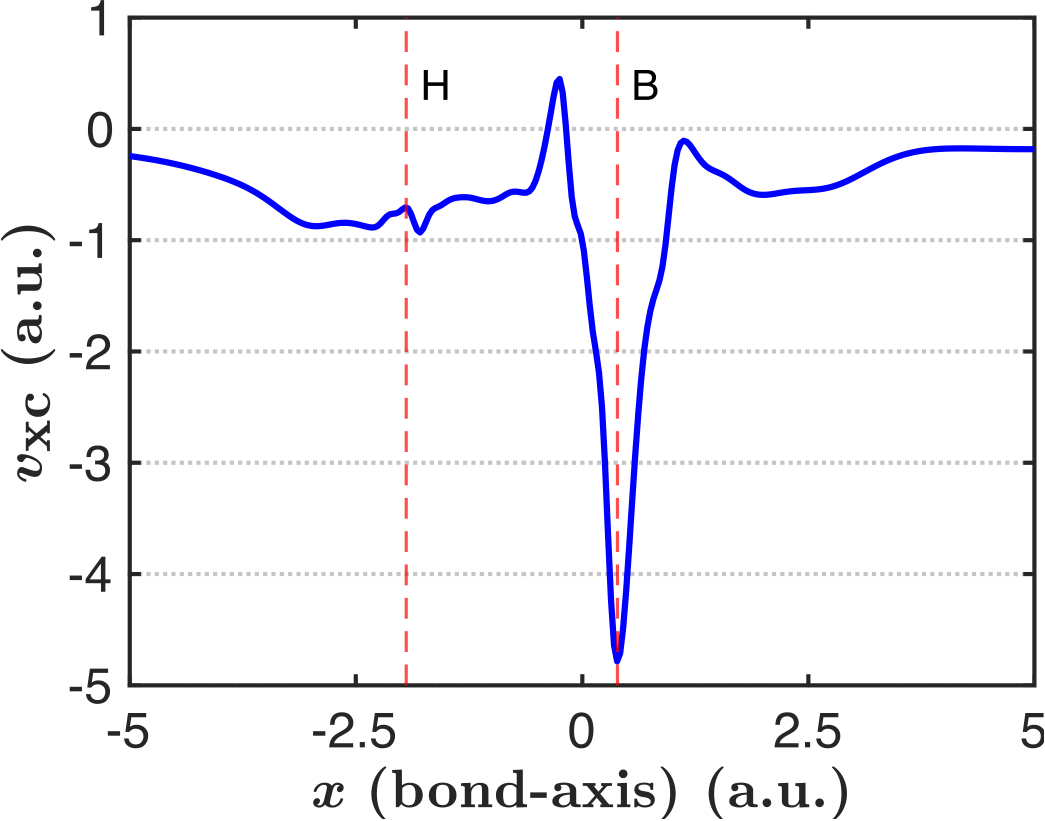}
     \caption{BH molecule}
     \label{fig:BH}
 \end{subfigure}
 \label{fig:Ne_BH}
\caption{Exact exchange-correlation potential for: (a) Ne atom and (b) BH molecule.}
\end{figure}

\noindent We, first, consider the Ne atom and the BH molecules where the target density ($\rhod$) are obtained using HBCI with the polarized Gaussian basis with tight cores (cc-pCVQZ)~\cite{Pritchard2019}. Figures~\ref{fig:Ne} and~\ref{fig:BH} present the exact XC potential for Ne and BH, respectively. As evident, \invDFT generates smooth potential for both the systems devoid of any finite basis artifacts. 

\begin{figure} 
    \centering
    \includegraphics[width=0.5\linewidth]{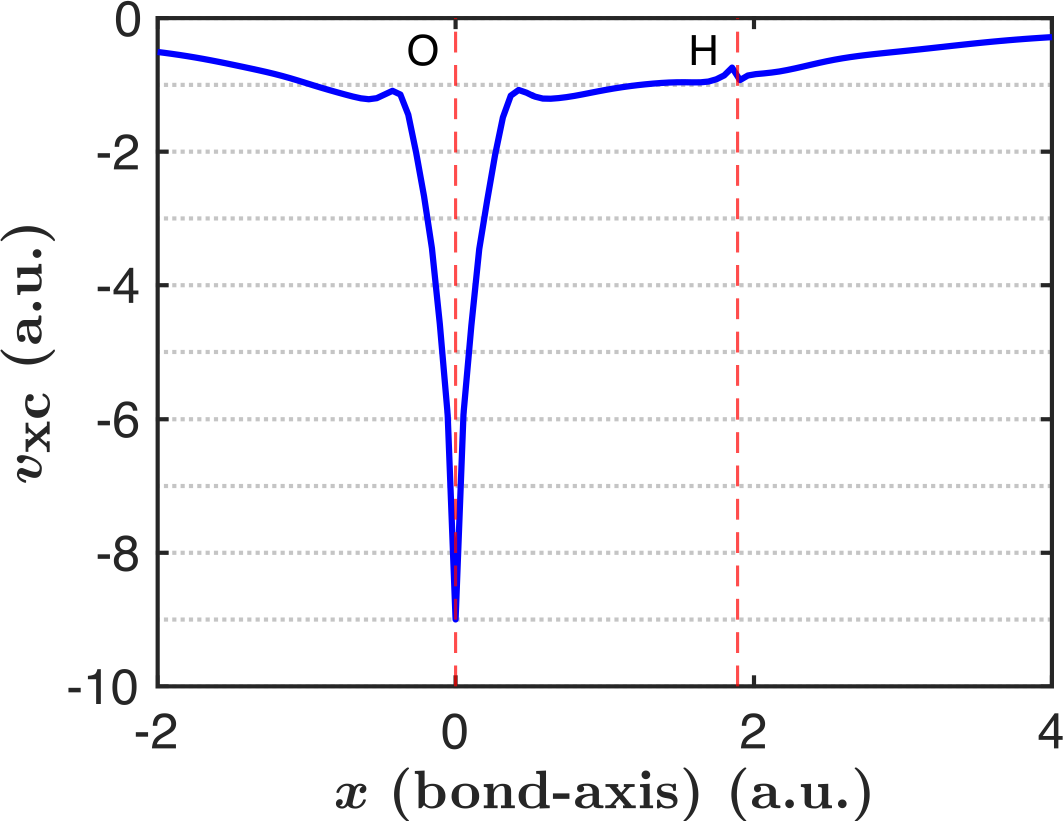}
    \caption{Exact exchange-correlation potential for H$_2$O along the O-H bond.}
    \label{fig:H2O}
\end{figure}

\begin{figure}[htbp!]
    \centering
    \includegraphics[scale=0.4]{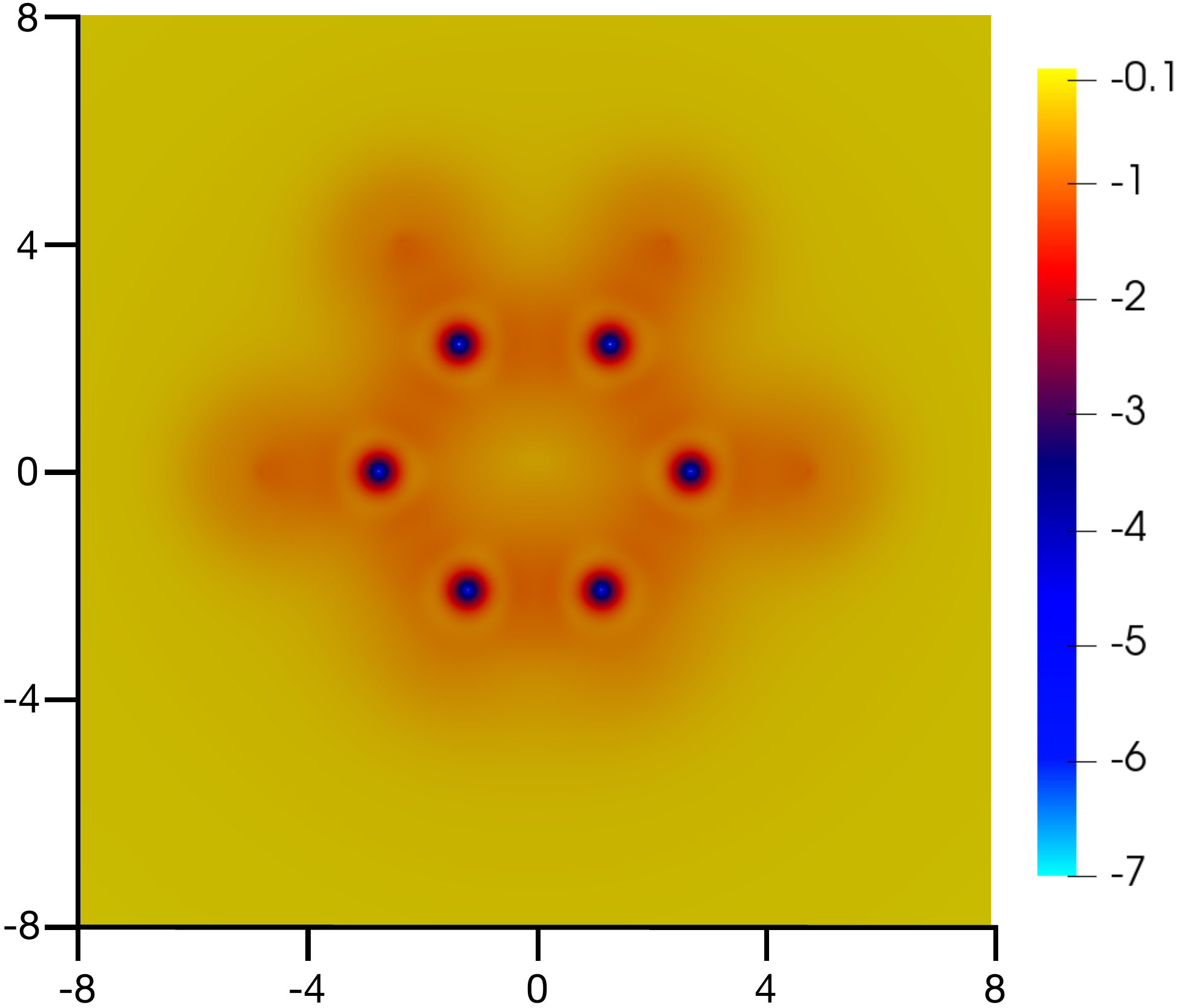}
    \caption{Exact exchange-correlation potential for ortho-benzyne (C$_6$H$_4$), presented on the plane of the molecule.}
    \label{fig:benzyne}
\end{figure}

\noindent Next, we consider two comparatively larger systems, namely, the water molecule (H$_2$O) and the  ortho-benzyne radical (C$_6$H$_4$). The target densities are obtained using iFCI with the aug-pcSseg-4 and cc-pVTZ Gaussian basis for H$_2$O and C$_6$H$_4$, respectively. The ortho-benzyne radical is strongly correlated species that has served as a stringent test for accurate wavefunction theories~\cite{Zimmerman2017b}. Figures~\ref{fig:H2O} and~\ref{fig:benzyne} depict the exact XC potential for H$_2$O and C$_6$H$_4$, respectively. Once again, \invDFT computes smooth potentials without any basis set artifacts. Both the potentials exhibit the intershell structure around the heavier atom---hump slightly away from the O in H$_2$O and the yellow rings around C atoms in C$_6$H$_4$. We also obtained excellent agreement between the KS highest occupied molecular orbital (HOMO) eigenvalue ($\epsH$) and the negative of the ionization potential ($I$), as  mandated by the Koopmans' theorem~\cite{Perdew1997}. Specifically, for H$_2$O, we have $\epsH=-0.452$ Ha, against $-I=-0.454$ Ha. Similarly, for C$_6$H$_4$, we have $\epsH=-0.354$ Ha against $\-I=-0.355$ Ha. We emphasize that the Koopmans' theorem is a stringent test for any inverse DFT, requiring tight accuracy in the low density region. Thus, meeting the Koopmans' theorem to good accuracy underlines the accuracy of \invDFT. 

\begin{figure}
 \medskip
 \begin{subfigure}{0.49\textwidth}
     \includegraphics[width=\textwidth]{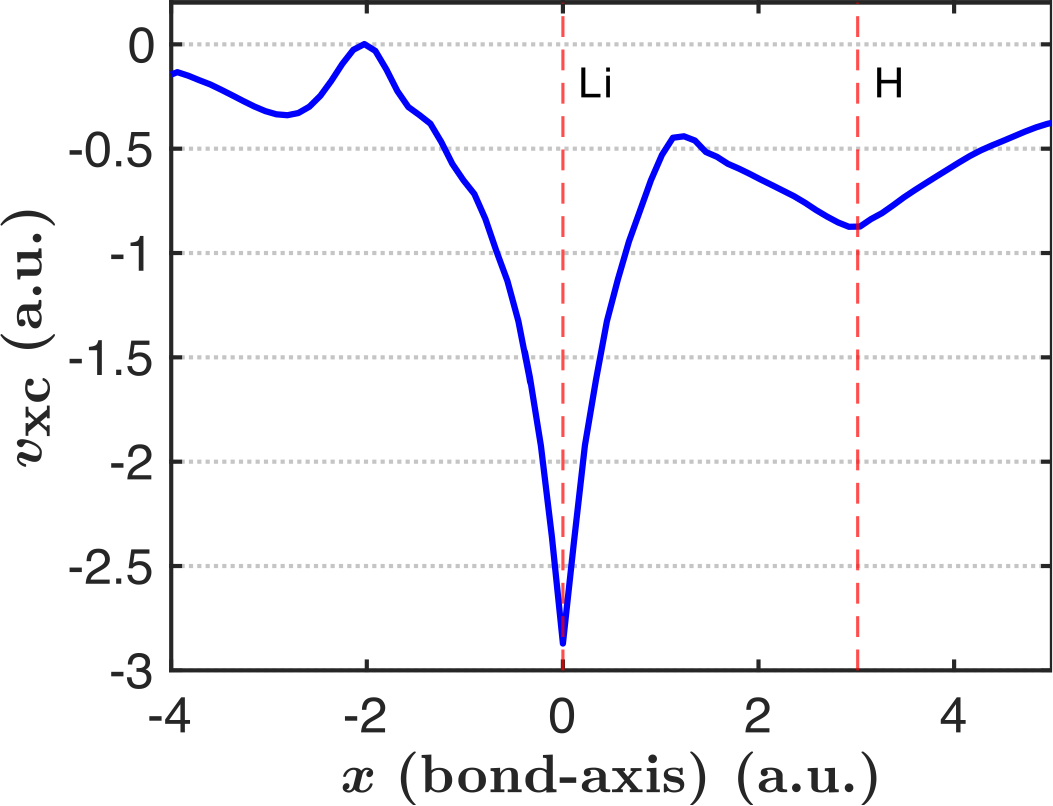}
     \caption{LiH molecule}
     \label{fig:LiH}
 \end{subfigure}
 \hfill
 \begin{subfigure}{0.49\textwidth}
     \includegraphics[width=\textwidth]{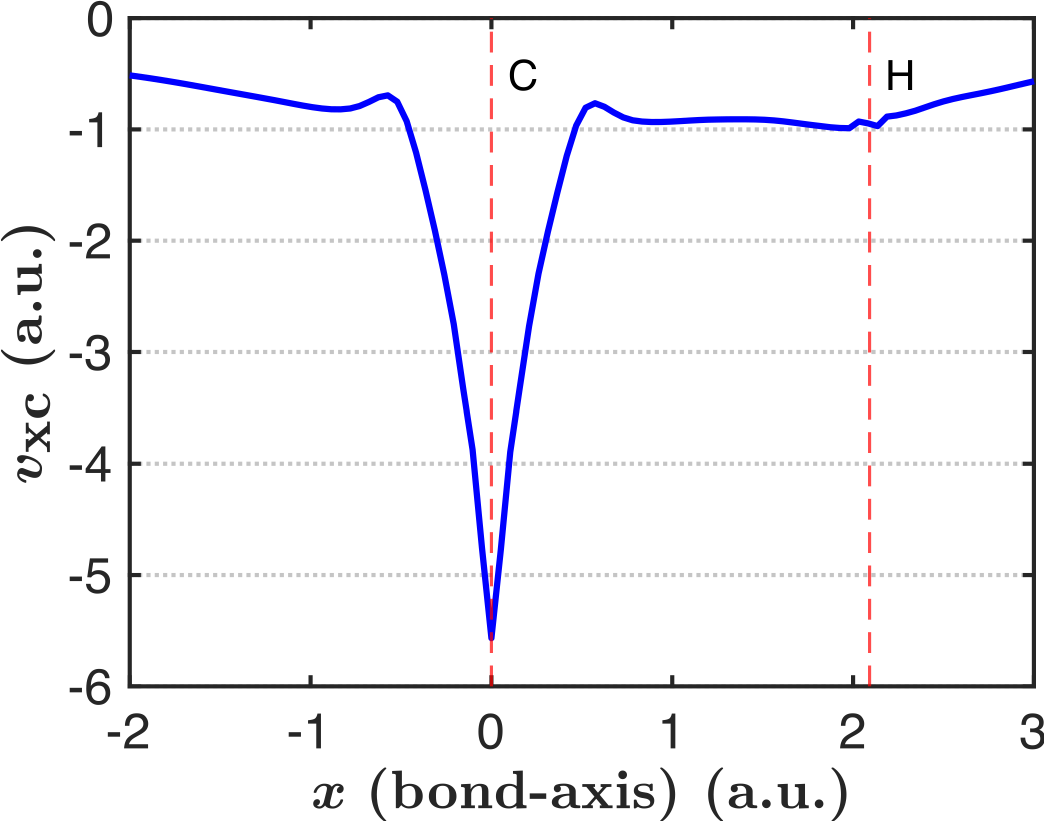}
     \caption{Singlet CH$_2$ molecule}
     \label{fig:CH2}
 \end{subfigure}
 \caption{Exact exchange-correlation potential for: (a) LiH molecule and (b) singlet CH$_2$ molecule (along C-H bond). }
\end{figure}

\noindent We now consider two systems---LiH molecule and singlet CH$_2$ radial---with Slater densities. Specifically, the target densities are obtained using HBCI with the QZ4P Slater basis developed by Van Lenthe and Baerends~\cite{Van2003}. The CH$_2$ radical is a strongly correlated system, forming another challenging benchmark for \invDFT. Figures~\ref{fig:LiH} and ~\ref{fig:CH2} present the exact XC potentials for LiH and CH$_2$, respectively. As with previous cases, we obtain smooth potentials as well as the intershell structure around the heavier atom.

\subsection{GPU Acceleration and Strong scaling}
\begin{figure}[htbp!]
    \centering
    \includegraphics[scale=0.3]{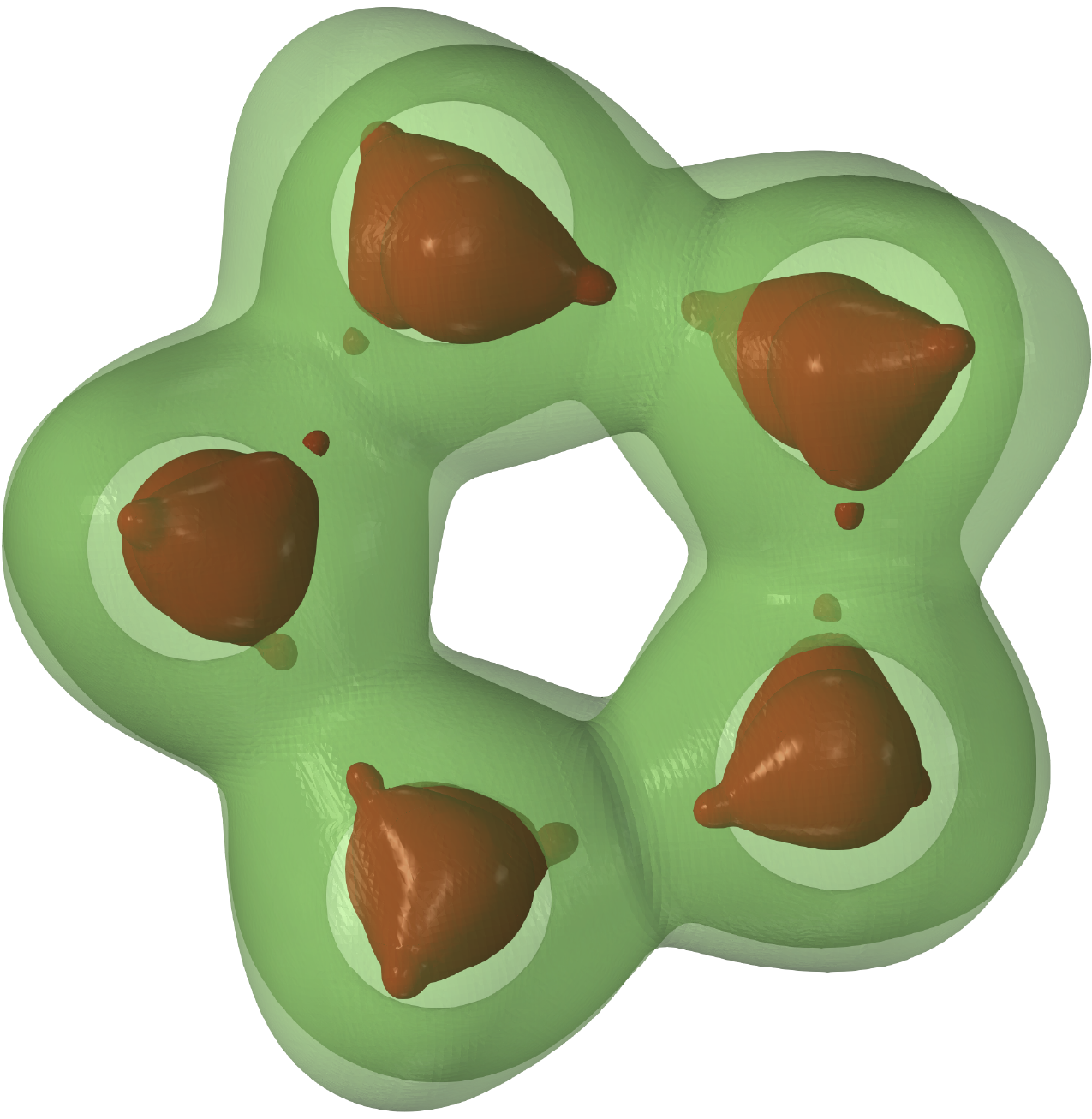}
    \caption{Exact exchange-correlation potential for water decamer ($(\text{H}_2\text{O})_{10}$). The green and brown contours correspond to values of -0.2 and -0.8 a.u., respectively.}
    \label{fig:H2O_decamer_2}
\end{figure}

\begin{figure}[htbp!]
    \centering
    \includegraphics[scale=0.4]{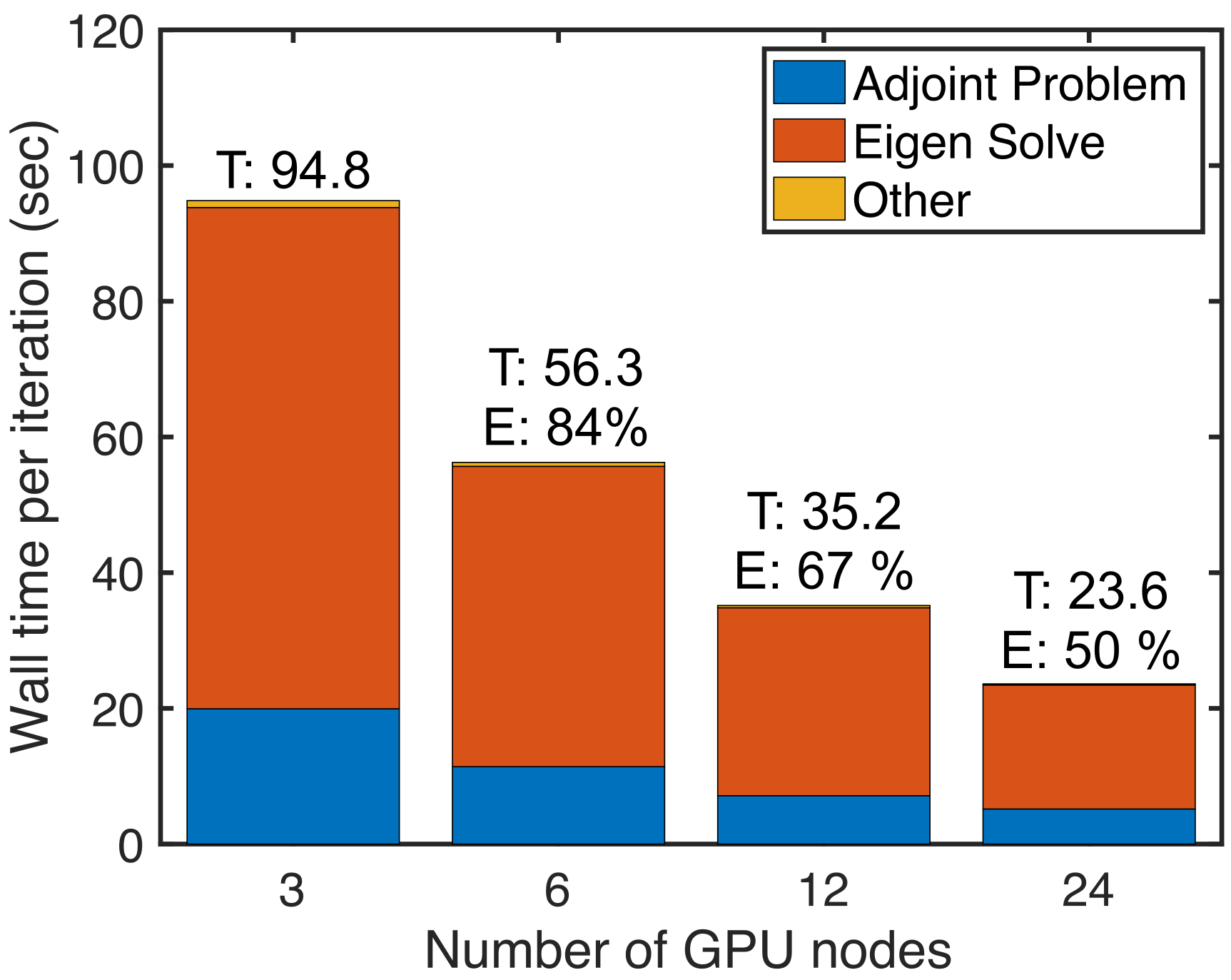}
    \caption{Strong scaling of \invDFT for ortho-benzyne (C$_6$H$_4$) using Perlmutter GPU nodes. Each node contains 4 GPUs. T denotes the walltime per BFGS iteration (in seconds) and E denotes the parallel efficiency.}
    \label{fig:benzyne_scaling}
\end{figure}

\begin{figure}[htbp!]
    \centering
    \includegraphics[scale=0.4]{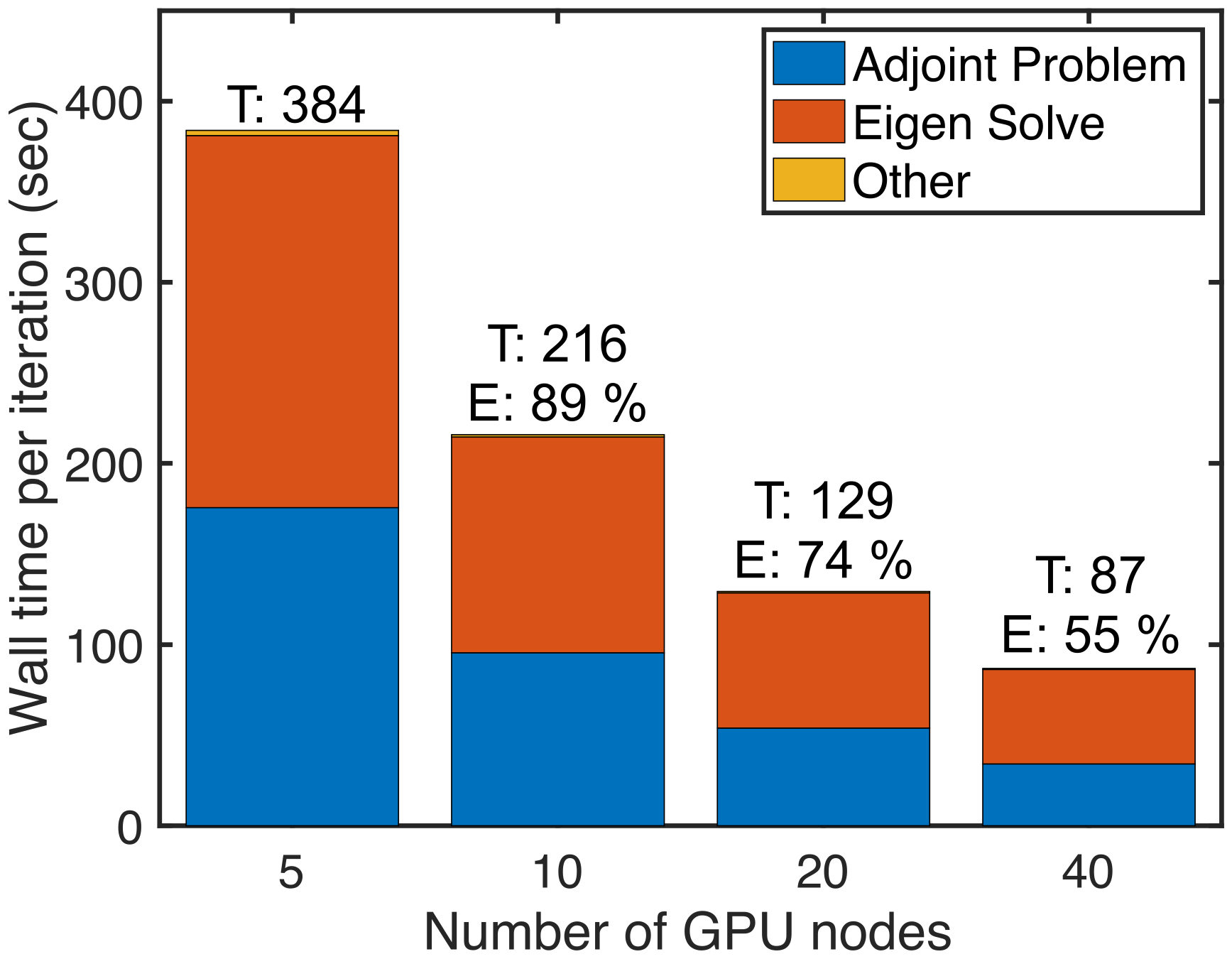}
    \caption{Strong scaling of \invDFT for water decamer ($(\text{H}_2\text{O})_{10}$) using Frontier GPU nodes. Each node contains 8 GPUs. T denotes the walltime per BFGS iteration (in seconds) and E denotes the parallel efficiency.}
    \label{fig:water_decamer_scaling}
\end{figure}

We demonstrate CPU-GPU speedup as well as the strong scaling of \invDFT using ortho-benzyne (C$_6$H$_4$) (40 electrons) and water decamer ($(\text{H}_2\text{O})_{10}$) (100 electrons) as two large benchmark system (large by inverse DFT standards). For ortho-benzyne we use the same iFCI density as reported in Sec.~\ref{sec:exactXC}. For the water decamer obtaining accurate groundstate densities via CI is difficult, owing to its high computational cost. Thus, for water decamer, for the purpose of demonstrating the strong scaling, we use the groundstate density obtained using the SCAN0 hybrid functional~\cite{Hui2016} as a proxy to the exact density. We remark that finding the exact XC potential for the SCAN0 density is non-trivial, given that it is an orbital-dependent functional. For both the systems, we discretize the KS orbitals and the adjoint functions using an adaptive sixth-order finite-element mesh containing 2.6 million and 7.2 million basis functions for ortho-benzyne and water decamer, respectively. Figure~\ref{fig:H2O_decamer_2} shows the XC potential for water decamer corresponding to the SCAN0 density. We attain a 14$\times$ and 9$\times$ CPU-GPU speedup for ortho-benzyne and water decamer, respectively,  on a node-to-node basis, highlighting efficient GPU acceleration of key computational kernels in \invDFT. Further, for both systems, we study the strong scaling for up to 8$\times$ the initial number of GPU nodes used. Fig.~\ref{fig:benzyne_scaling} and Fig.~\ref{fig:water_decamer_scaling} present the strong scaling for ortho-benzyne and water decamer, respectively. As evident, for both systems, we attain a good parallel scalability of 50\% or higher at 8$\times$ the minimum GPU nodes, i.e., at 24 nodes for ortho-benzyne and 40 nodes for water decamer. Further, all parts of \invDFT scale equally well. For ortho-benzyne, we attain a walltime of 24 seconds per BFGS iteration on 24 GPU nodes. Similarly, for water decamer, we attain a walltime of 87 seconds per BFGS iteration at 40 GPU nodes. Considering $\sim$500 BFGS iterations, for systems $<50$ electrons, one can perform an inverse DFT calculation within 4 hours, and, for systems with 50-100 electrons in within 12 hours of walltime. This presents a $40-50\times$ speedup over our initial implementation~\cite{Kanungo2019}, attributed to the efficient solver strategies and HPC considerations incorporated in \invDFT (see Sec.~\ref{sec:solver} and Sec.~\ref{sec:hpc}). 

\section{Conclusions}
\noindent In this work, we have developed \invDFT as a massively parallel CPU-GPU tool for accurate inverse DFT calculations to find the exact XC potential from a target density. Inverse DFT, for long, remained a challenging problem, owing to numerical artifacts arising from the use of atomic orbital basis (Gaussian or Slater) employed in previous efforts. These difficulties have also curtailed the development of efficient software packages for inverse DFT. To date, to the best of our knowledge, only three software packages for inverse DFT exist. However, all three packages employ Gaussian basis, which is prone to non-unique solutions and/or spurious oscillations in the XC potentials. \invDFT alleviates the numerical issues in inverse DFT through a combination of systematically convergent finite-element basis as well as enforcement of appropriate boundary conditions. Together, these advances render the problem well-posed and result in smooth and accurate XC potentials. Currently, \invDFT supports inverse DFT within the spin-restricted formalism and for finite systems (atoms, molecules). It allows for inverse DFT on both Gaussian and Slater densities, in a format and manner which can be easily fed from widely used quantum chemistry codes. 

\noindent For greater computational speed, we employed the Chebyshev filtering based subspace iteration and preconditioned MINRES method to solve the Kohn-Sham eigenvalue problem  and the adjoint problem, respectively. We also incorporated various HPC advances into \invDFT to boost its performance. This includes, judicious batching over vectors, within the eigenvalue and the adjoint problem, to simultaneously maximize arithmetic intensity as well as minimize data-access and data-movement costs. Further, to gainfully exploit the fine-grained parallelism in modern CPU-GPU architectures, we leveraged on \DFTFE to recast the global sparse-dense matrix products to several simultaneous finite-element level dense-dense matrix products. Given that \invDFT involves two different finite-element meshes, we have also developed an efficient method for transfer of solution fields between incompatibly partitioned meshes, which allows for greater scalability. 

\noindent We validated the accuracy of \invDFT by reconstructing the XC potentials from LDA densities. We also presented the exact XC potentials for various atoms and molecules, spanning both weak and strongly correlated systems, using accurate full configuration interaction densities. Wherever possible, we demonstrated the XC potentials to satisfy the Koopmans' theorem, which constitutes a stringent test of the accuracy of the potential. We also assessed the performance of \invDFT in terms of its CPU-GPU speed and its strong scaling. We attained a 9-15$\times$ speedup on GPUs over CPUs, in a node-to-node comparison. Further, we attained good parallel scalability up to 40 GPU nodes for a system with 100 electrons. Overall, \invDFT allows for routine inverse DFT calculation ($<50$ electrons) within 4 hours of walltime---a marked $40-50\times$ improvement over our previous inverse DFT implementation.

\noindent Inverse DFT is vital link between quantum many-body methods and DFT. By finding the exact XC potential, it helps elucidate the nature of electronic correlations. It also serves as a powerful tool to study and assess approximate functionals in terms of their XC potentials to shed more light into their accuracy, structure and deficiencies~\cite{Kanungo2021}. More importantly, it provides for crucial data that can be used to model better XC functionals, either via machine-learning~\cite{Kanungo2024b, Schmidt2019} or other empirical methods. Although inverse DFT methods have progressed recently, an efficient and scalable software package had been lacking. \invDFT fills that gap. While currently \invDFT supports only spin-restricted formulation of inverse DFT for finite systems (atoms and molecules), our ongoing efforts will extend it to spin-unrestricted formalism as well as to extended systems (solids). In future, we also envisage the use of an enriched finite element basis~\cite{Kanungo2017, Rufus2021} to further accelerate inverse DFT calculations.

\section*{Acknowledgment}
We thank Paul M. Zimmerman and Jeffrey Hatch for providing accurate configuration interaction (CI) based groundstate densities used in this work. We thank Sambit Das for help with the integration of \invDFT with \DFTFE. This work is supported by Department of Energy, Office of Science, through grant no. DE-SC0022241. This study used resources of the NERSC Center, a DOE Office of Science User Facility using NERSC award BES-ERCAP0034270. This study also used resources of the Oak Ridge Leadership Computing Facility, which is a DOE Office of Science User Facility supported under Contract DE-AC05-00OR22725.

\bibliography{ref}
\bibliographystyle{unsrt}
\end{document}